\documentclass[12pt,preprint]{aastex}
\usepackage{longtable}
\usepackage{pdflscape}	
\usepackage{multirow}

\begin{document}

\title{The first light curve modeling and orbital period change investigation of nine contact binaries around the short period cut-off}

\author{Li Kai\altaffilmark{1,2}, Kim, Chun-Hwey\altaffilmark{3}, Xia, Qi-Qi\altaffilmark{1}, Michel, Raul\altaffilmark{4}, Hu, Shao-Ming\altaffilmark{1}, Gao, Xing\altaffilmark{5}, Guo, Di-Fu\altaffilmark{1}, Chen, Xu\altaffilmark{1}}

\altaffiltext{1}{Shandong Key Laboratory of Optical Astronomy and Solar-Terrestrial Environment, School of Space Science and Physics, Institute of Space Sciences, Shandong University, Weihai, Shandong, 264209, China (e-mail: kaili@sdu.edu.cn (Li, K.))}
\altaffiltext{2}{Institute for Astrophysics, Chungbuk National University, Cheongju 28644, Republic of Korea}
\altaffiltext{3}{Department of Astronomy and Space Science, Chungbuk National University, Cheongju 361-763, Korea}
\altaffiltext{4}{Observatorio Astron\'{o}mico Nacional, Instituto de Astronom\'{\i}a, Universidad Nacional Aut\'{o}noma de M\'{e}xico, Apartado Postal 877, Ensenada, B.C. 22830, M\'{e}xico}
\altaffiltext{5}{Xinjiang Astronomical Observatory, 150 Science 1-Street, Urumqi 830011, China}

\begin{abstract}
In this paper, we present the first light curve synthesis and orbital period change analysis of nine contact binaries around the short period limit. It is found that all these systems are W-subtype contact binaries. One of them is a medium contact system while the others are shallow contact ones. Four of them manifest obvious O'Connell effect explained by a dark spot or hot spot on one of the component stars. Third light was detected in three systems. By investigating orbital period variations, we found that four of the targets display a secular period decrease while the others exhibit a long-term period increase. The secular period decrease is more likely caused by angular momentum loss while the long-term period increase is due to mass transfer from the less massive component to the more massive one. Based on the statistic of 19 ultrashort period contact binaries with known orbital period changes, we found that seven of them display long-term decrease (three of them also exhibit cyclic variations), ten of them manifest long-term increase while two of them only show cyclic variation and that most of them are shallow contact binaries supporting the long timescale angular momentum loss theory suggested by Stepien. For the three deep contact systems, we found that they are probably triple systems. The tertiary companion plays an essential role during their formation and evolution.

\end{abstract}

\keywords{stars: binaries: close ---
         stars: binaries: eclipsing ---
         stars: evolution ---
         stars: individual }

\section{Introduction}
Contact binaries are generally composed of two Roche lobe overfilling F, G, or K type stars. Although they are very common in the universe, their formation, evolution and ultimate fate are still poorly understood (e.g., Bradstreet \& Guinan 1994; Qian 2003; Yakut \& Eggleton 2005; Stepien 2006; Eggleton2012). Rucinski (1992) firstly noticed a 0.22 day short period cut-off by using the data of contact binaries in the 4th edition of the General Catalogue of Variable Stars (GCVS). More recently, Drake et al. (2014a) presented the period distribution of the potential contact binaries in the Catalina Data Release 1 while Qian et al. (2017) showed the distribution of orbital period of contact binaries based on the Variable Star Index (VSX) and the Large Sky Area Multi-Object Fiber Spectroscopic Telescope (LAMOST) databases. Li et al. (2019) have carried out statistics on the period distribution of contact binaries according to the VSX and GCVS databases and most results of the worldwide photometric surveys. All these period distribution studies show that an obviously sharp decline at about 0.22 days can be identified.

Since the discovery of this short period limit, researchers have tried to solve this issue. Rucinski (1992) claimed that the fully convective limit can possibly explain the short period cut-off. Stepien (2006, 2011) suggested that the timescale of the angular momentum loss (AML) is too long to form extremely short period contact binaries. Jiang et al. (2012) proposed that the low mass limit at about 0.63 $M_\odot$ of the primary component causes the short period limit. Very recently, Li et al. (2019) claimed that a third companion is very important during the formation of some ultrashort period contact binaries (USPCBs\nolinebreak\footnotemark[1] \footnotetext[1]{contact binaries with orbital periods shorter than 0.23 days.}) by removing the AML from the central eclipsing star. At present, the short period limit issue is still an open question, more contact binaries around the short period cut-off should be observed and analyzed. In this paper, we present the first photometric investigations and orbital period analyses of nine USPCBs. Table 1 shows the basic information of these nine targets.

\begin{table*}
\tiny
\begin{center}
\caption{The information of the studied stars}
\begin{tabular}{lccccccccc}
\hline
Star	                          & hereafter    &	RA	        &  Dec	      	&Period (days) &HJD$_0$ &V (mag)&Amplitude&References        \\
&name&&&&2450000+&&&\\\hline
1SWASP J003033.05+574347.6	    & J003033	      & 00 30 33.11 &	+57 43 47.3	  &0.226618	     &7675.2292(2)&  14.89	 &0.68     & (1), (2)\\
CRTS J014418.3+190625	          & J014418	      & 01 44 18.30 &	+19 06 25.9	  &0.217372	     &8112.9738(2)&  15.97	 &0.67     & (3)      \\
1SWASP J031700.67+190839.6	    & J031700	      & 03 17 00.69 &	+19 08 39.4	  &0.225652	     &8051.3043(1)&  14.06	 &0.48     & (2), (3) \\
CRTS J074350.9+451620	       & J074350	      & 07 43 50.87 &	+45 16 20.6	  &0.227324	     &8482.3208(2)&  15.95	 &0.40     & (3)     \\
1SWASP J104942.44+141021.5	    & J104942	      & 10 49 42.44 &	+14 10 21.5	  &0.229799	     &8162.1155(4)&  14.31	 &0.25     & (2), (3) \\
CRTS J130945.0+371627	       & J130945	      & 13 09 45.12 &	+37 16 26.9	  &0.211132	     &8541.0006(2)&  16.27	 &0.46     & (3) \\
CRTS J145224.5+011522	          & J145224	      & 14 52 24.55 &	+01 15 21.7	  &0.196014	     &7732.1679(2)&  15.05	 &0.58     & (3)         \\
1SWASP J161858.05+261303.5$^a$	& J161857	      & 16 18 57.86 &	+26 13 38.5	  &0.228781	     &8289.3108(3)&  14.39	 &0.21$^b$  & (2)    \\
CRTS J224015.4+184738    	     & J224015	      & 22 40 15.45 &	+18 47 38.4	  &0.218802	     &8362.0972(2)&  15.45	 &0.43     & (3)      \\  \hline
\end{tabular}
\end{center}
References: (1) Norton et al. 2011; (2) Lohr et al. 2013b; (3) Drake et al. 2014b\\
$^a$ The coordinates of this target were determined by Lohr et al. (2013b), but when we processed the observational data, we found that the position of the variable star is $\alpha_{2000.0}=16^h18^m57.86^s$, $\delta_{2000.0}=26^{\circ} 13^{\prime}38.5^{\prime\prime}$. The short name was changed to J161857, and the corrected coordinates are shown in the third and forth columns of this table.\\
$^b$ The variability amplitude of this star was roughly determined with the public data of SuperWASP.\\
\end{table*}

\section{Observations and Data Reduction}
The observations of these USPCBs were carried out with five meter-class telescopes from 2016 to 2019. These five telescopes are the Weihai Observatory 1.0-m telescope of Shandong University (WHOT, Hu et al. 2014), the 85 cm telescope at the Xinglong Station of National Astronomical Observatories (NAOs85cm) in China, the 60cm Ningbo Bureau of Education and Xinjiang Observatory Telescope (NEXT), and the 84cm and 2.12m telescopes of the San Pedro Martir Observatory (SPM84cm and SPM2.12m). The PIXIS 2048B CCD was used for WHOT, the Andor DZ936 CCD was used for NAOs85cm, the FLI PL23042 CCD was used for NEXT, the Spectral Instruments 1 CCD was used for SPM84cm, and the Marconi 4 CCD was used for SPM2.12m, resulting in the field of views as $12^{'}\times12^{'}$, $32^{'}\times32^{'}$, $22^{'}\times22^{'}$, $7.6^{'}\times7.6^{'}$, and $6^{'}\times6^{'}$ for the five telescopes, respectively. During the observations, the standard Johnson-Cousins filters were employed. Observing details of all stars including the observing date, exposure time, observational errors, the comparison and check stars, and the telescope are listed in Table 2. The standard aperture photometry reduction technique was applied to process all observing data by using IRAF routines\nolinebreak\footnotemark[1] \footnotetext[2]{IRAF is distributed by the National Optical Astronomy Observatories, which is operated by the Association of Universities for Research in Astronomy Inc., under contract to the National Science Foundation.}. After the data reduction, the differential magnitudes between the target and a comparison star were derived. In order to confirm that the brightness of the comparison star was constant, a check star was also introduced. Figure 1 shows the light curves of J031700 as an example. As seen in the lower panel of Figure 1, the different magnitudes between the comparison and check stars are flat, thus indicating that the brightness of the comparison star is constant.

\renewcommand\arraystretch{1.3}
\begin{table*}
\tiny
\caption{Observing details of the nine targets}
\begin{tabular}{p{0.8cm}p{2.2cm}p{1.9cm}p{2.7cm}p{2.3cm}p{2.3cm}p{1cm}}
\hline
Target     &Observing Date	    &Exposure time	            &Observational errors 	       & Comparison Star         & Check Star              & Telescope \\
           &                    &      (s)                   &      (mag)                       & 2MASS                   & 2MASS                &\\\hline
J003033	   &2016 Oct 13	        &$R_c$90 $I_c$70            &$R_c$0.016 $I_c$0.013	           & J00302874+5741007 & J00302526+5743428 & WHOT      \\
J014418	   &2017 Dec 25	        &$R_c$80  $I_c$50 	        &$R_c$0.015 $I_c$0.018	           & J01442029+1906584 & J01441092+1905281 & WHOT      \\
J031700	   &2017 Oct 24	        &$V$60  $R_c$25  $I_c$15 	  &$V$0.008 $R_c$0.007 $I_c$0.008    & J03165659+1909246 & J03170401+1912053 & WHOT        \\
J074350	   &2018 Dec 29	        &$R_c$85  $I_c$80 	        &$R_c$0.015 $I_c$0.017             & J07434571+4515593 & J07434393+4515443 & NAOs85cm        \\
J104942	   &2018 Feb 12	        &$V$80  $R_c$40  $I_c$25 	  &$V$0.012 $R_c$0.010 $I_c$0.010	   & J10494282+1412327 & J10492540+1410439 & WHOT        \\
J130945	   &2018 Apr 6, 8, 9	  &$V$20  $R_c$12  $I_c$12    &$V$0.009 $R_c$0.009 $I_c$0.009    & J13093799+3715199 & J13092906+3715120 & SPM2.12m        \\
           &2019 Jan 27, Feb 26	&$V$120  $R_c$50  $I_c$30  	&$V$0.011 $R_c$0.010 $I_c$0.010    & J13093799+3715199 & J13092906+3715120 & SPM84cm        \\
J145224	   &2017 May 13	        &$R_c$80  $I_c$40 	        &$R_c$0.016 $I_c$0.016	           & J14522682+0112500 & J14522617+0114086 & WHOT      \\
J161857	   &2018 Jun 06, 19	    &$R_c$40  $I_c$70 	        &$R_c$0.013 $I_c$0.012	           & J16191196+2612405 & J16191091+2609106 & NEXT      \\
J224015	   &2018 Aug 31	        &$R_c$100  $I_c$60 	        &$R_c$0.009 $I_c$0.010	           & J22404066+1846467 & J22401952+1848000 & WHOT      \\\hline
\end{tabular}
\end{table*}

\begin{figure*}
\begin{center}
\includegraphics[width=0.6\textwidth]{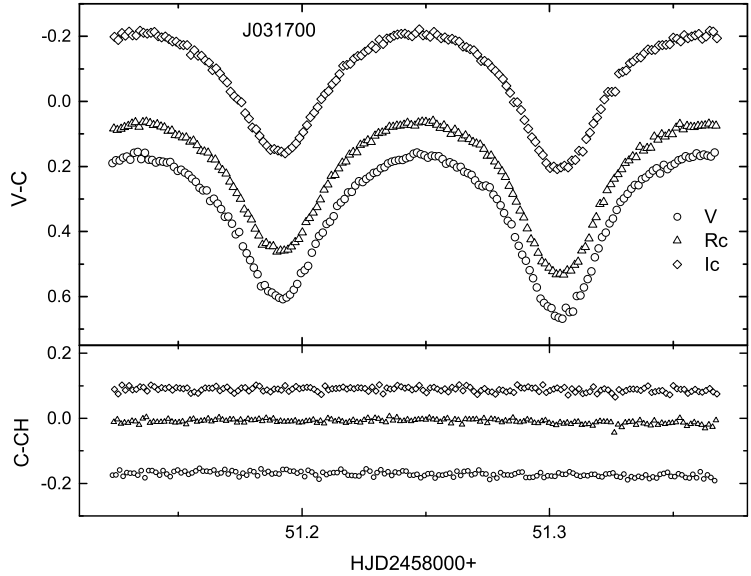}
\caption{The $VR_cI_c$ differential light curves of J031700 observed on October 24, 2017. The upper panel shows the differential magnitudes between the target and the comparison star, while the lower panel plots those between the comparison and check stars. Circles, triangles, and diamonds represent $V$, $R_c$, and $I_c$ measurements, respectively.}
\end{center}
\end{figure*}

\section{Photometric Solutions}
The light curves of the nine stars were all analyzed by using the 2013 version of Wilson-Devinney (W-D) code (Wilson \& Devinney 1971; Wilson 1979, 1990, 1994). First, we have to estimate reliable effective temperatures for these stars. The temperatures can be estimated by using their color indices. All stars lie in distances of several hundred pc, up to more than one kpc. In addition, some of them have small galactic latitudes. Hence, the color indices are very likely reddened. Therefore, the interstellar extinction for each star should be considered when estimating its temperature. The interstellar extinctions of different filters for each target were determined by using the method provided by Schlegel et al. (1998) based on the IRAS database\nolinebreak\footnotemark[1] \footnotetext[3]{https://irsa.ipac.caltech.edu/applications/DUST/}. For example, the values of E(B-V) for J003033, J014418, J031700, J074350, J104942, J130945, J145224, J161857, and J224015 were determined to be 0.469 mag, 0.055 mag, 0.148 mag, 0.048 mag, 0.033 mag, 0.013 mag, 0.044 mag, 0.055 mag, and 0.056 mag, respectively.
The temperatures derived by different dereddened color indices and spectroscopic observations are shown in Table 3. T$_{BV}$ is the temperature due to the dereddened $B-V$ (APASS DR9, Henden et al. 2015) color index and Table 5 of Pecaut \& Mamajek (2013), T$_{JK}$ is the temperature due to the dereddened $J-K$ (2MASS, Cutri et al. 2003) color index and Table 5 of Pecaut \& Mamajek (2013), T$_{gi}$ is the temperature due to the dereddened $g-i$ (SDSS DR12, Alam et al. 2015) color index and Table 3 of Covey et al. (2007), and T$_{spec}$ is the temperature due to the spectroscopic observation by LAMOST (Luo et al. 2015) or Drake et al. (2014a). Gaia temperatures (Gaia Collaboration et al. 2018) were not used in our study because of irrespective of interstellar extinction.
The last column of Table 3 displays the average values of the temperatures derived by different ways, the corresponding errors represent the standard deviations. Actually, this average value is neither the primary's nor the secondary's, it should be a quadrature temperature of the two components. Therefore, when modeling the light curves of these systems, we set this temperature as the primary's (primary means the hotter component), $T_1=T_m$, and the final temperatures of the primary and secondary components were derived by using the following equations (Kjurkchieva et al. 2018):
\begin{eqnarray}
T_1&=&T_m+{c\Delta T\over c+1},  \\\nonumber
T_2&=&T_1-\Delta T,
\end{eqnarray}
where $\Delta T=T_1-T_2$ and $c=L_2/L_1$ can be determined by the adopted final photometric results of each target.
Since the temperatures of all the systems are less than 5200 K, the gravity darkening and bolometric albedo coefficients were fixed as $g_{1,2}=0.32$ and $A_{1,2} = 0.5$, and bolometric and bandpass limb-darkening coefficients were automatically determined from Van Hamme (1993) by adopting a logarithmic function (ld=-3).

Because all the nine targets were studied for the first time, a $q$-search method was used to determine their mass ratios. Since our light curves show that all the systems are contact binaries, mode 3 (contact mode) of the W-D code was used during the analysis. When searching for the mass ratio, some parameters were adjustable: the orbital inclination $i$; the effective temperature of star 2, $T_2$; the monochromatic luminosity of star 1, $L_1$; and the dimensionless potential ($\Omega=\Omega_1=\Omega_2$). The relationship between the resulting sum of weighted square deviations ($\Sigma W_i(O-C)^2_i$) and $q$ for each system is displayed in Figure 2. As seen in this figure, every system has a minimum value of $\Sigma$. This value was set as an initial value and an adjustable parameter, then new solutions were performed until a convergent solution was obtained. The determined parameters for each system are displayed in Table 4, and the corresponding synthetic light curves, solid black lines, are plotted in Figure 3. We found that some of the targets manifest asymmetric light curves, in these cases spot model was applied to fit their light curves (e.g., Qian et al. 2007; Lee et al. 2013; Zhou et al. 2016a). In addition, the search of a third light ($l_3$) was carried out for every system. The derived results are also shown in Table 4, and the corresponding theoretical light curves with spot and including third light, respectively, labeled with blue and red lines are displayed in Figure 3. We found that four stars exhibit spot activity and three stars reveal third light. We should state that when we searched for the third light, the spot parameters were fixed to quickly obtain a convergent solution. When we determined the final photometric results of all stars, the temperatures of their two components were calculated using Equation (1) and are listed in Table 5. The errors were derived by error propagate.

\begin{table}
\begin{center}
\caption{The temperatures of the nine targets determined by different methods}
\begin{tabular}{lccccc}
\hline
Target     & T$_{BV}$ (K)& T$_{JK}$ (K)&T$_{gi}$ (K)&T$_{spec}$ (K)   & $T_m$ (K) \\
\hline
J003033    &  5140     &  5130    &  -     & -     & 5135$\pm7  $  \\
J014418    &  4990     &  4480    &  4600  & -     & 4690$\pm270$  \\
J031700    &  5000     &  4610    &  4900  & -     & 4837$\pm200$  \\
J074350    &  -        &  3750    &  4700  & 4660$^a$  & 4370$\pm540$  \\
J104942    &  4930     &  3880    &  4700  & -     & 4503$\pm550$  \\
J130945    &  -        &  3500    &  -     & 3870$^a$  & 3685$\pm260$  \\
J145224    &  3700     &  3880    &  3850  & 3940$^b$  & 3843$\pm100$  \\
J161857    &  4940     &  4690    &  3800  & -     & 4477$\pm600$  \\
J224015    &  4690     &  4200    &  4650  & -     & 4513$\pm270$  \\
\hline
\end{tabular}
\end{center}
$^a$ These two temperatures were derived by the observations of LAMOST.\\
$^b$ This temperature was determined based on the spectroscopic observation of Drake et al. (2014a).
\end{table}

\begin{figure}\centering
\includegraphics[width=0.32\textwidth]{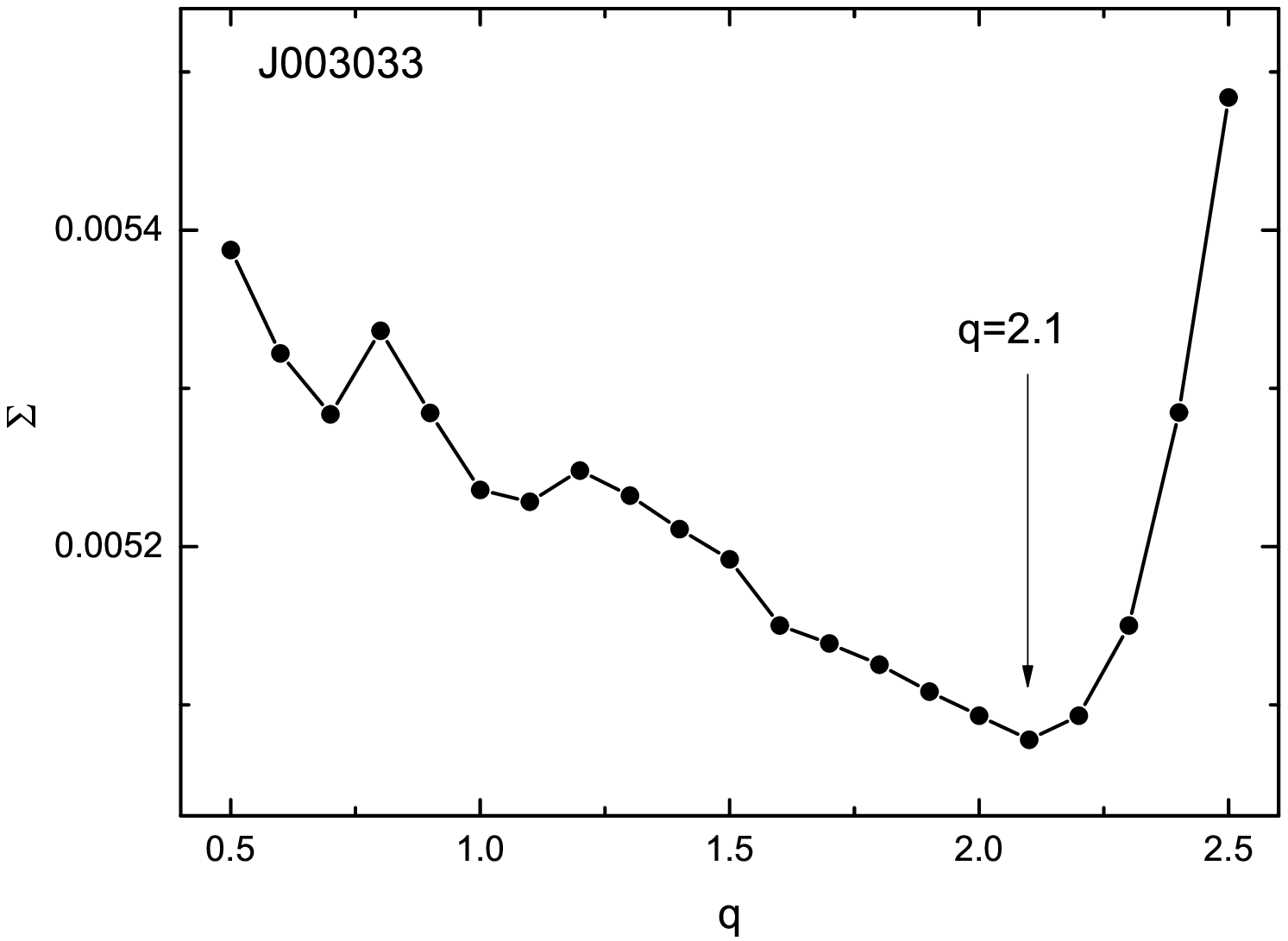}
\includegraphics[width=0.32\textwidth]{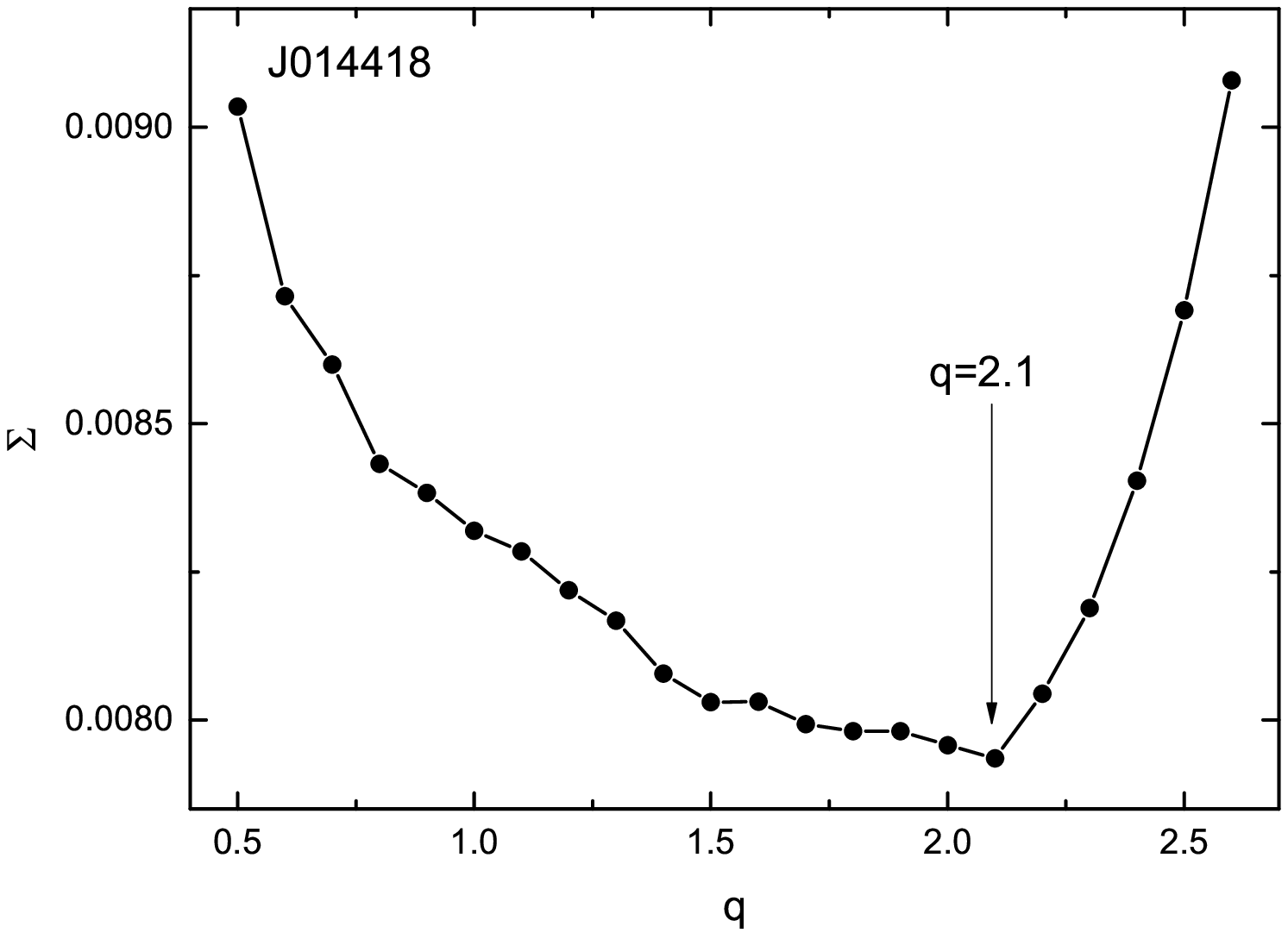}
\includegraphics[width=0.32\textwidth]{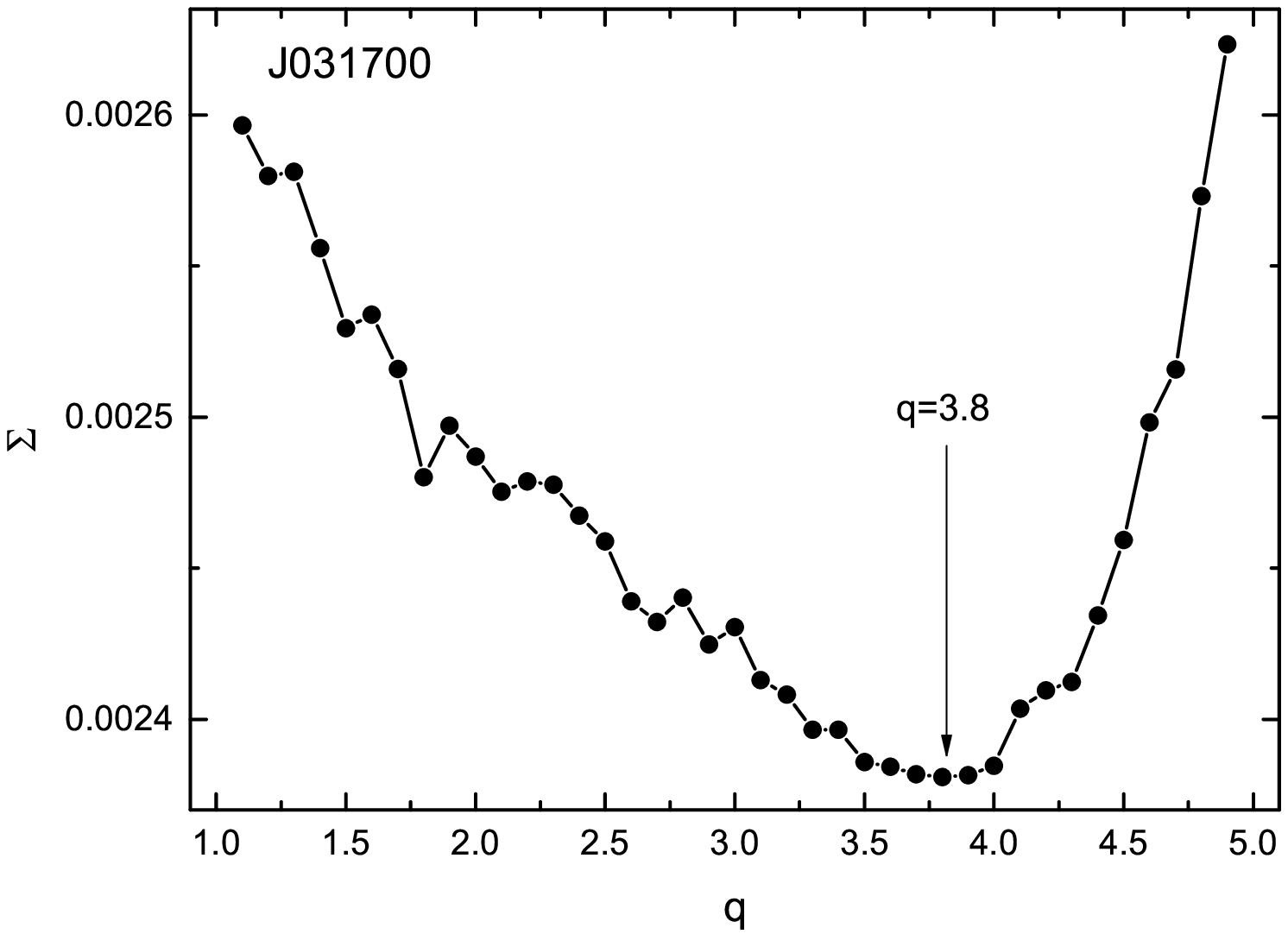}
\includegraphics[width=0.32\textwidth]{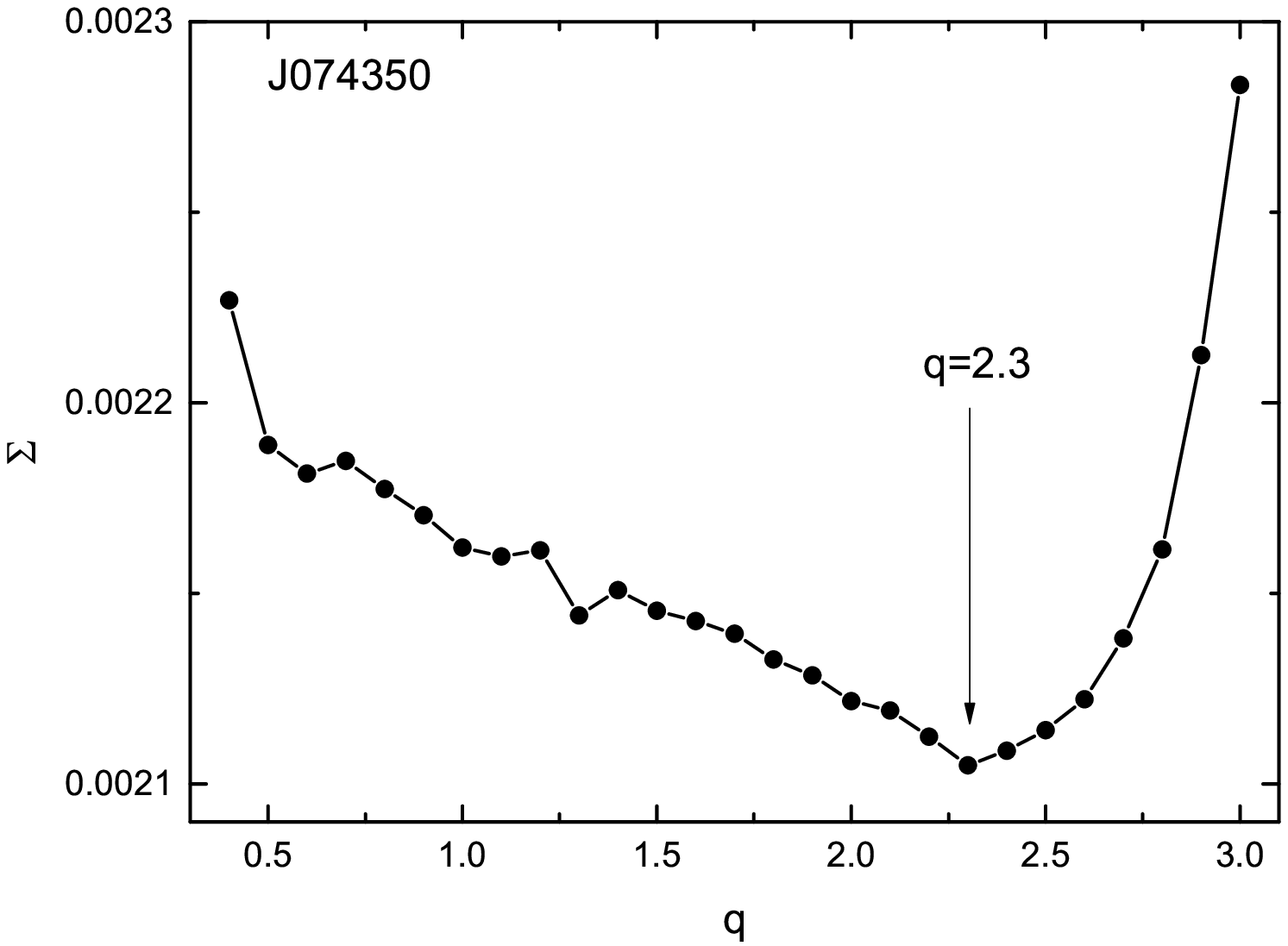}
\includegraphics[width=0.32\textwidth]{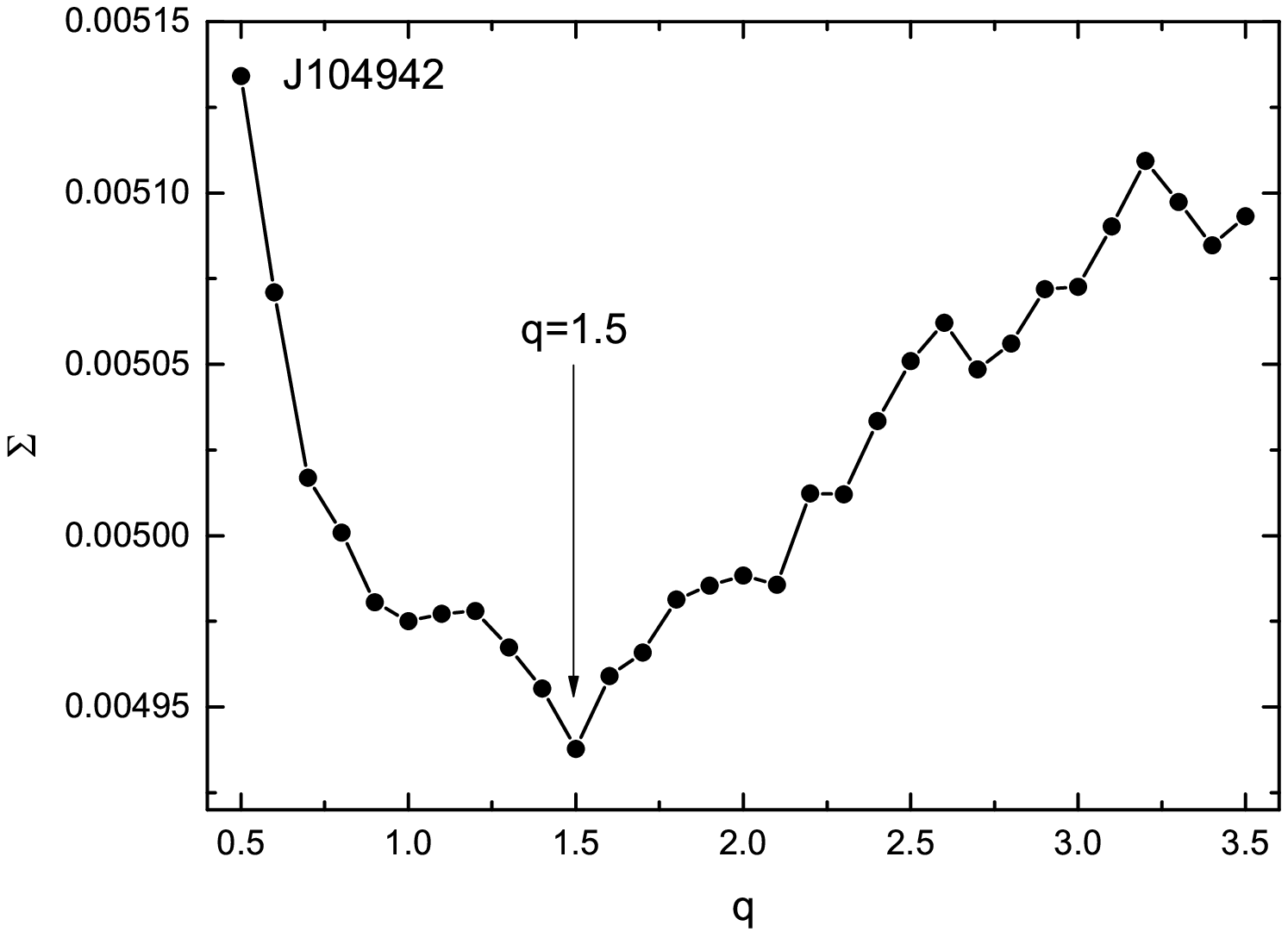}
\includegraphics[width=0.32\textwidth]{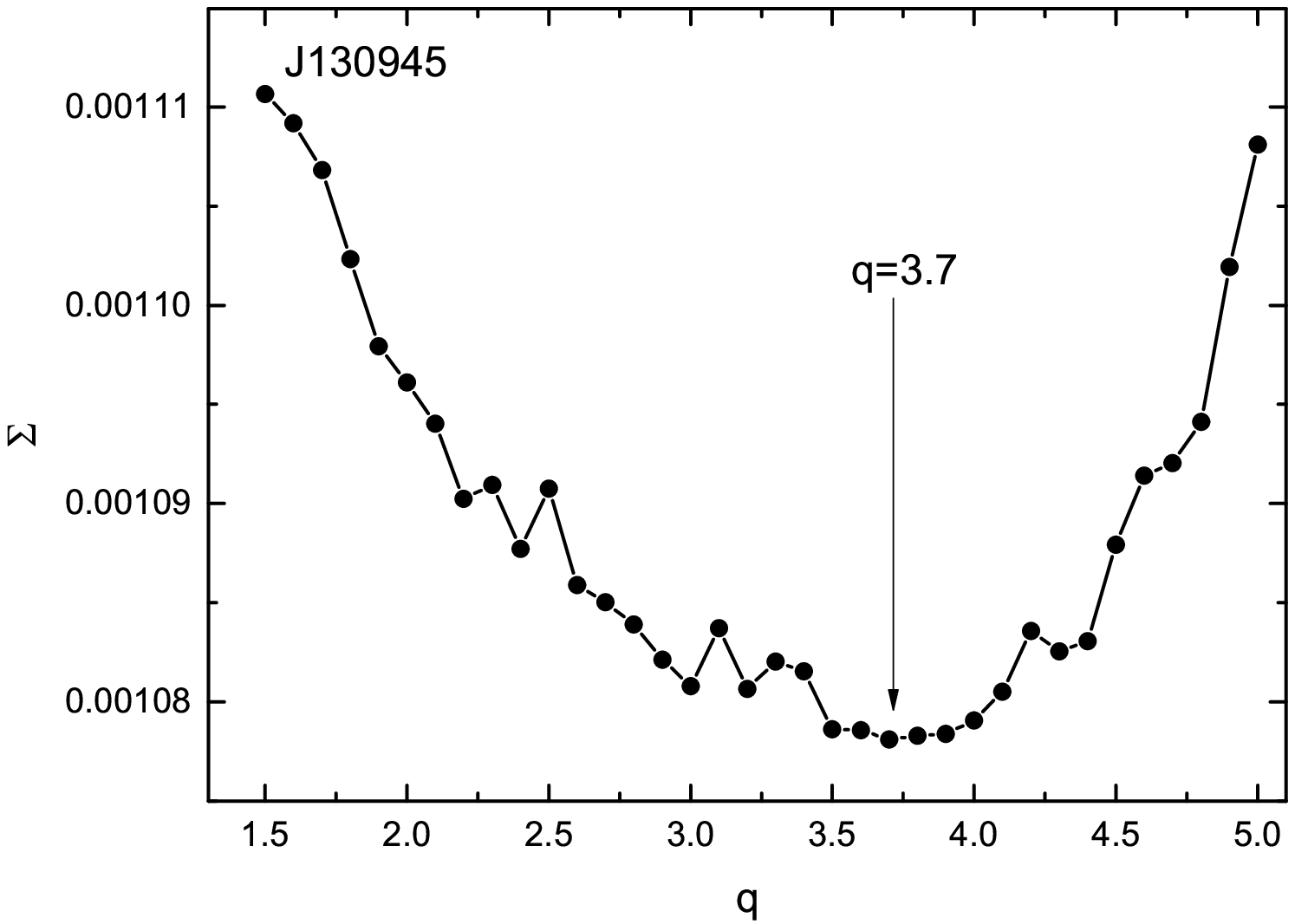}
\includegraphics[width=0.32\textwidth]{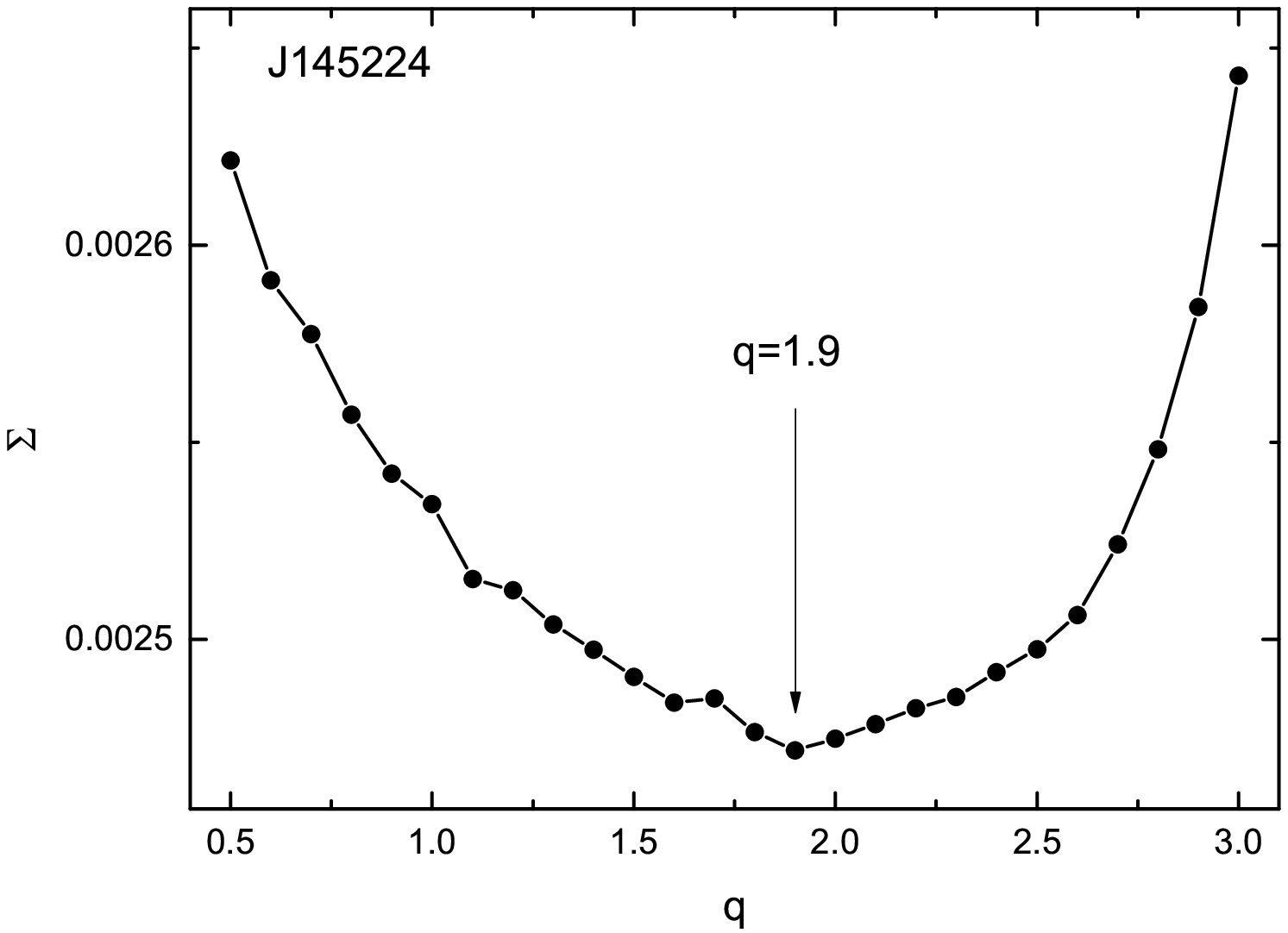}
\includegraphics[width=0.32\textwidth]{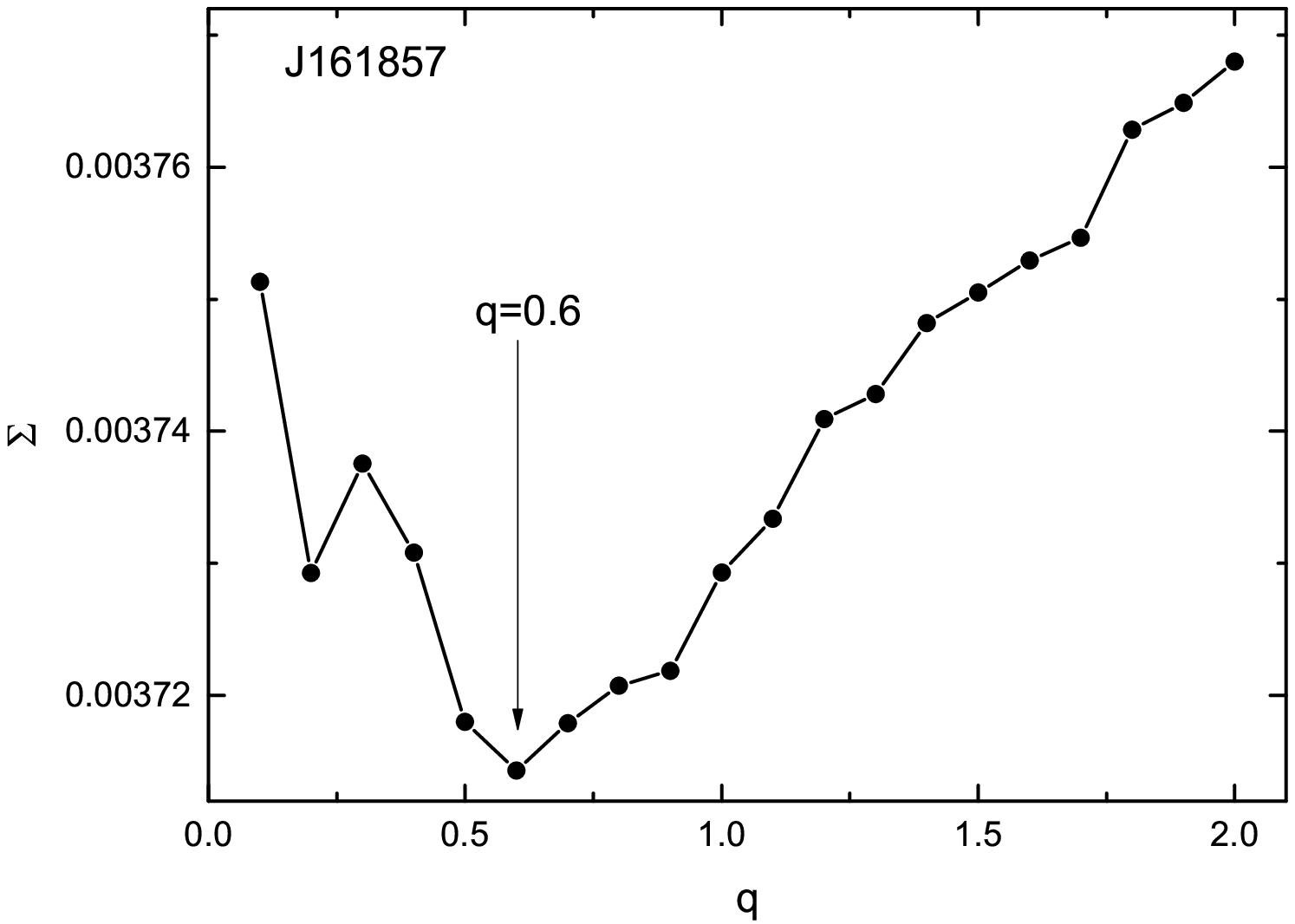}
\includegraphics[width=0.32\textwidth]{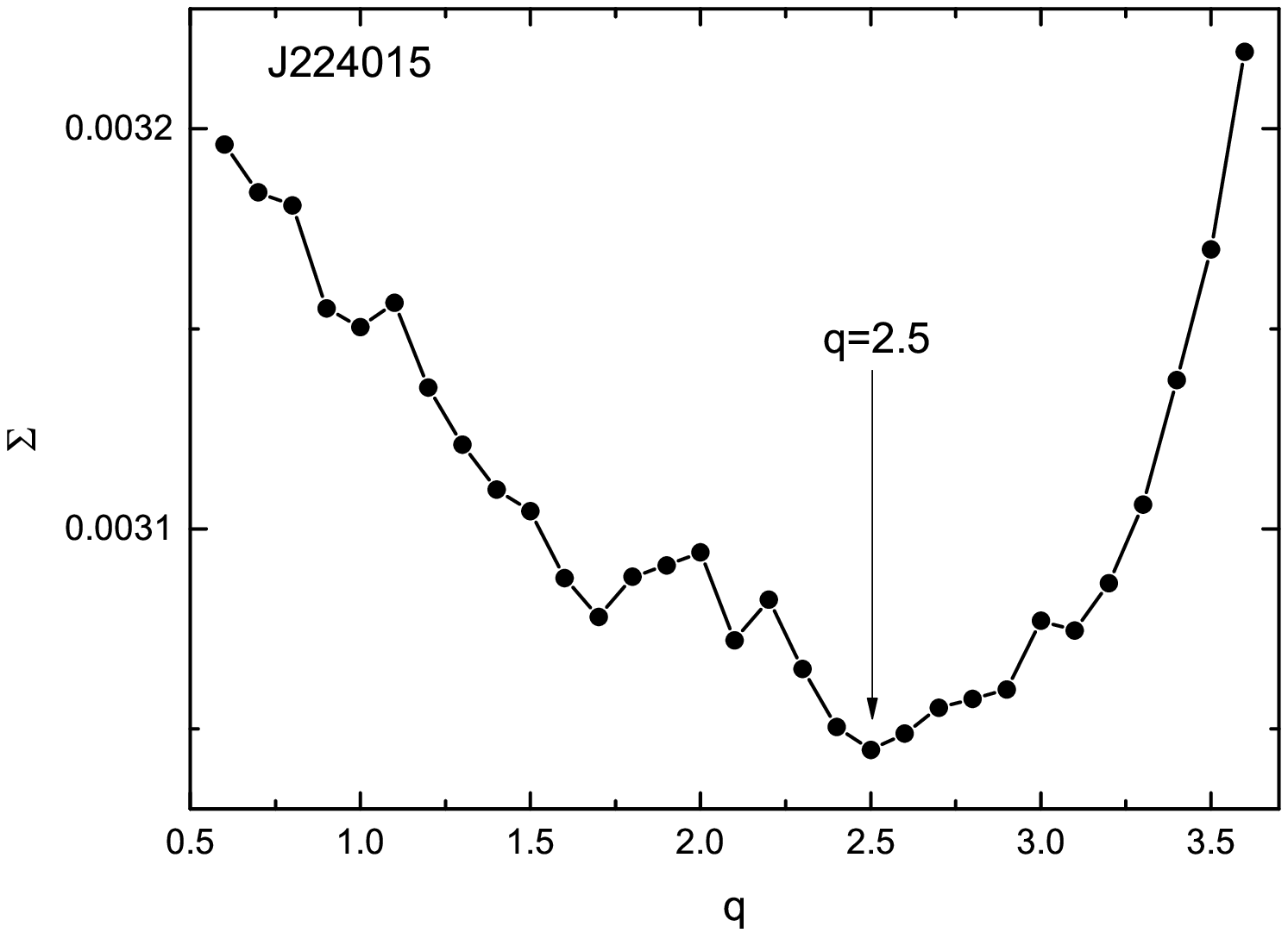}
\caption{ Plots of $\Sigma W_i(O-C)^2_i$ versus mass ratio $q$ for the nine stars.}
\end{figure}

\begin{figure}\centering
\includegraphics[width=0.32\textwidth]{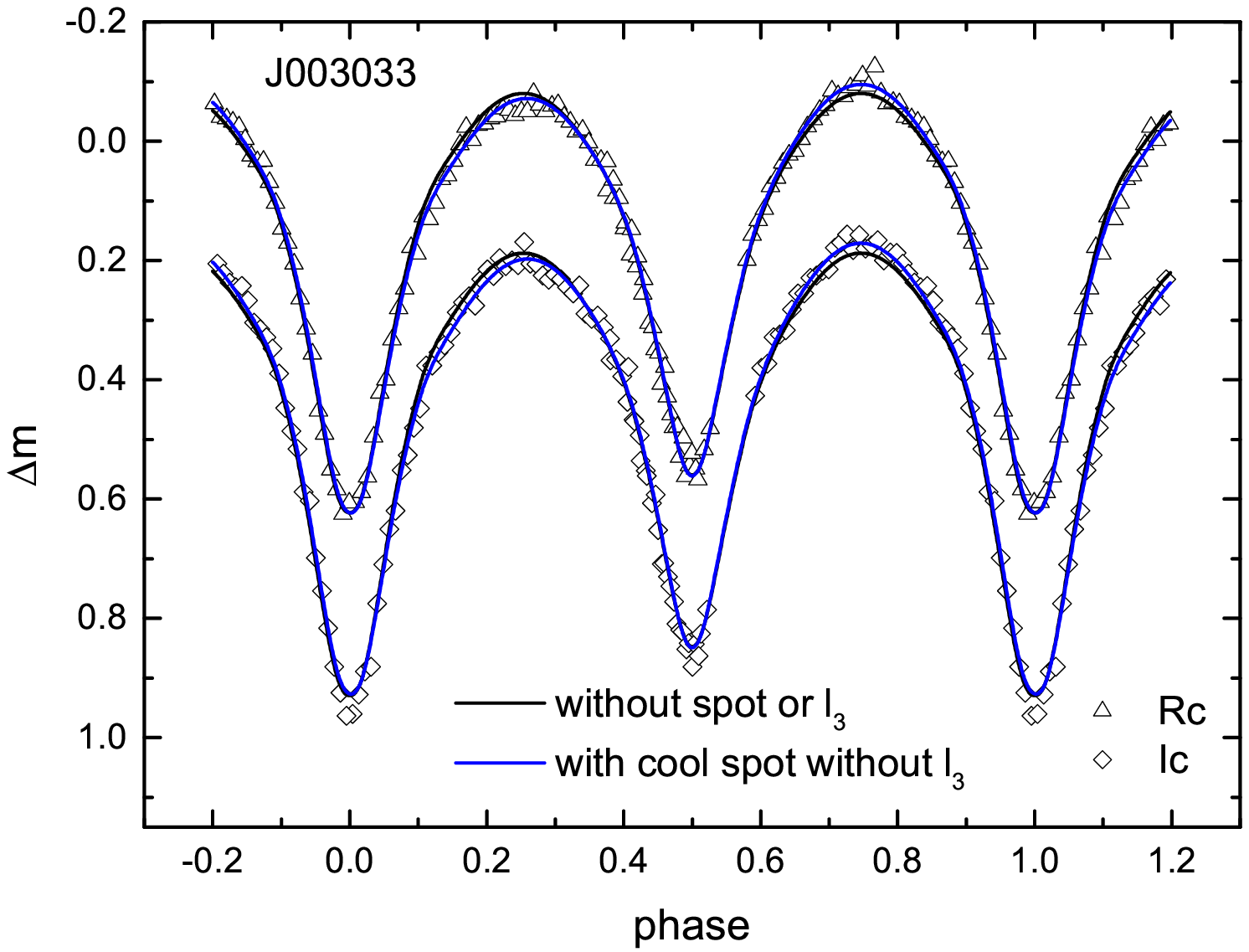}
\includegraphics[width=0.32\textwidth]{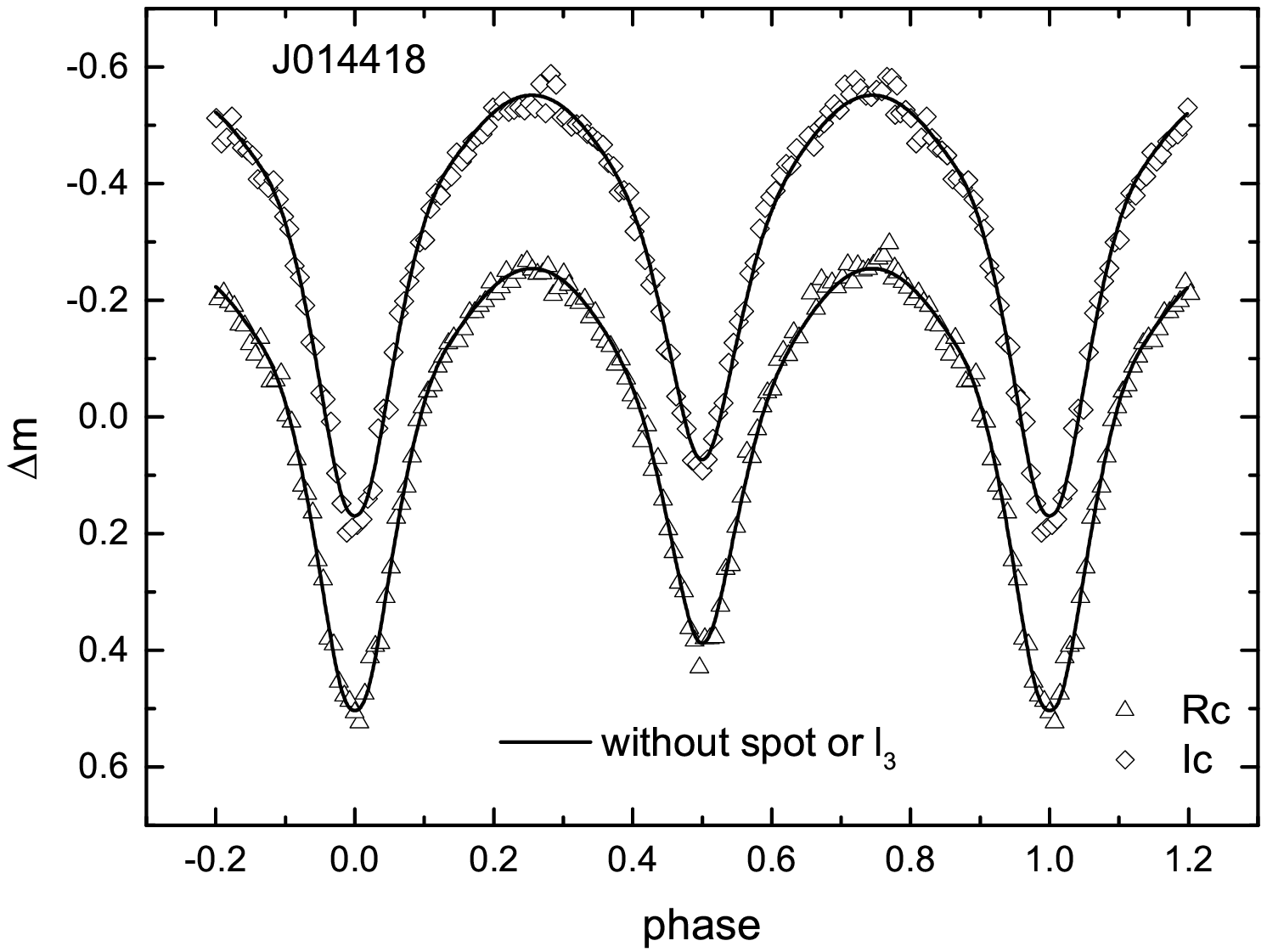}
\includegraphics[width=0.32\textwidth]{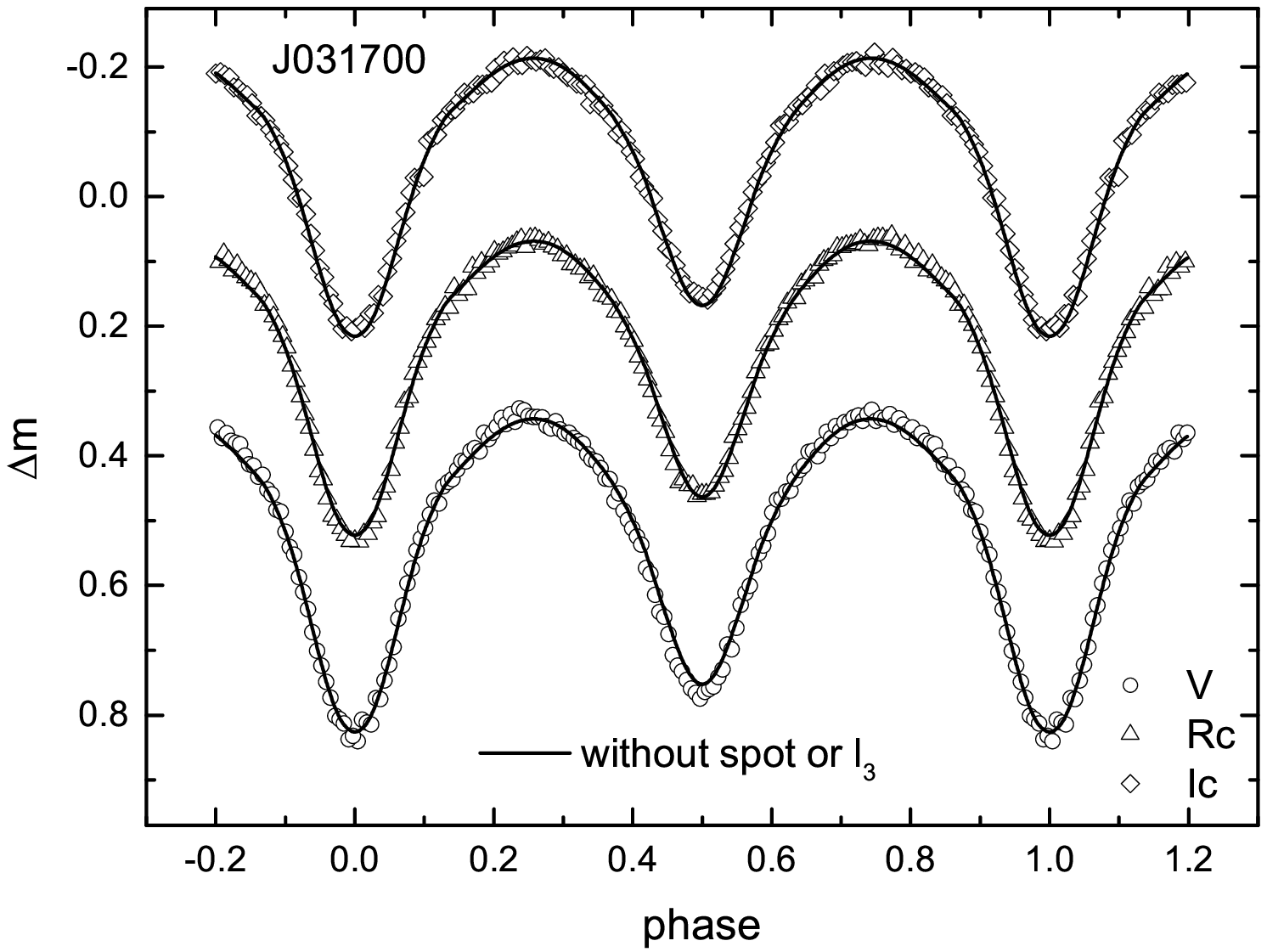}
\includegraphics[width=0.32\textwidth]{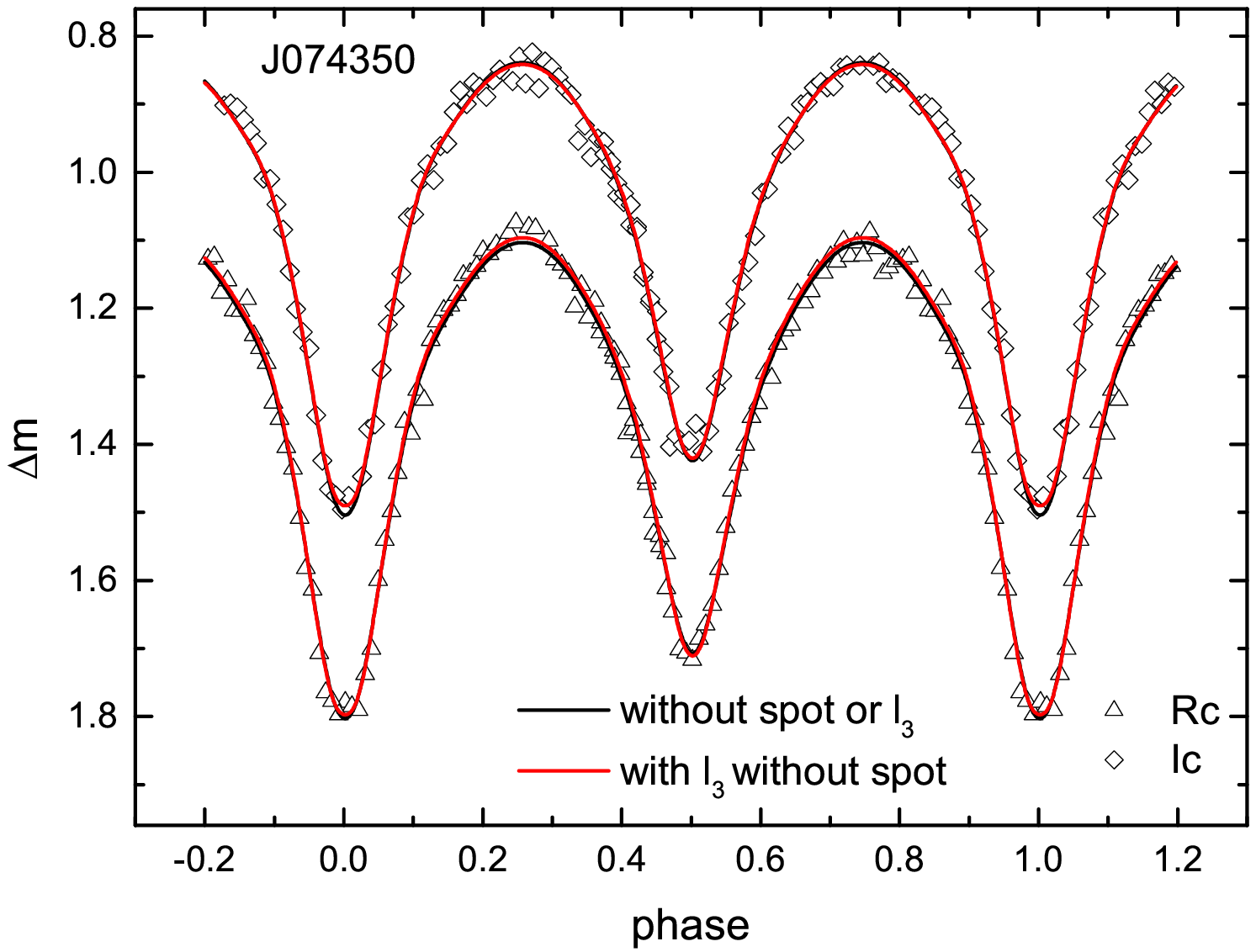}
\includegraphics[width=0.32\textwidth]{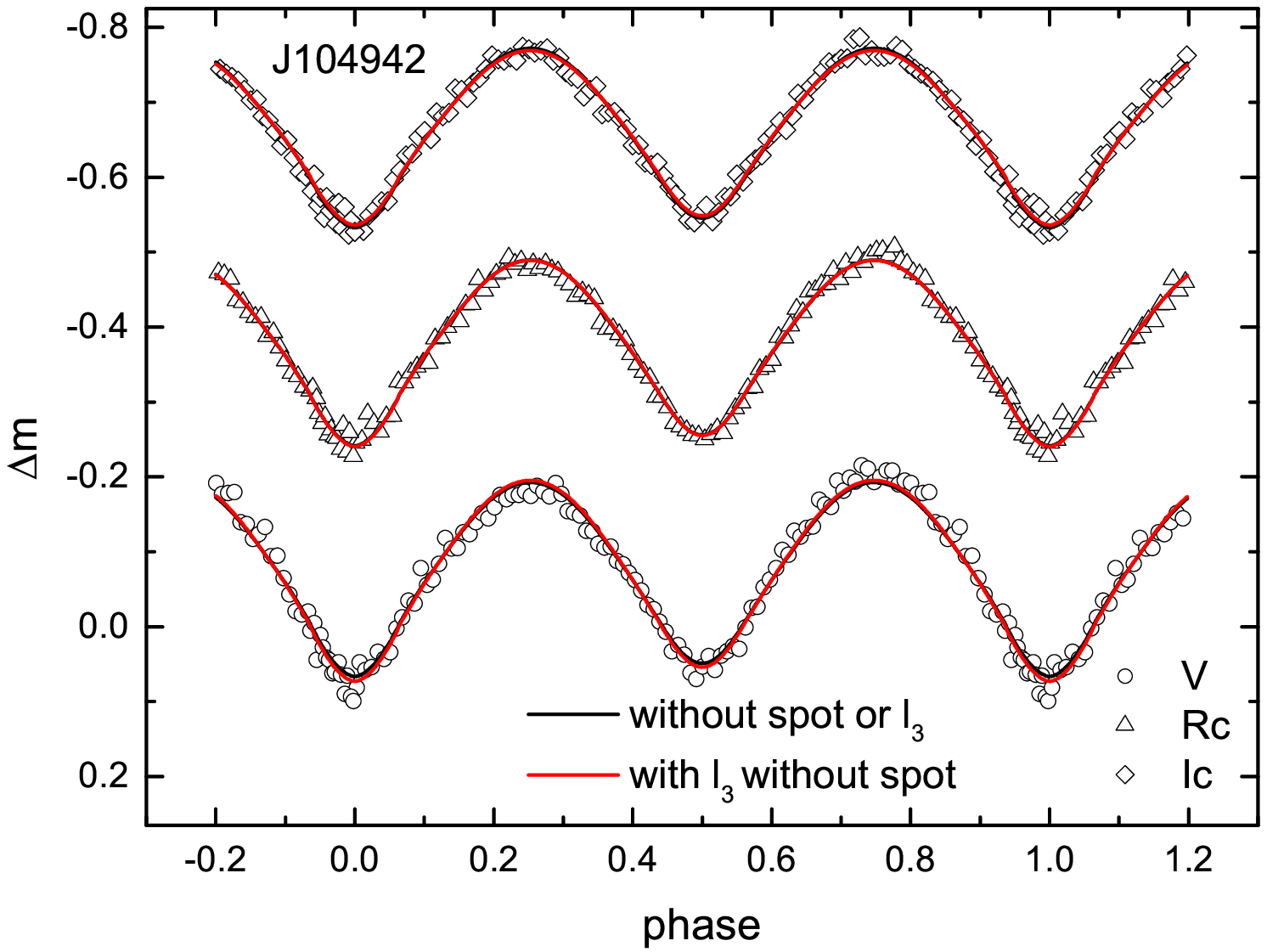}
\includegraphics[width=0.32\textwidth]{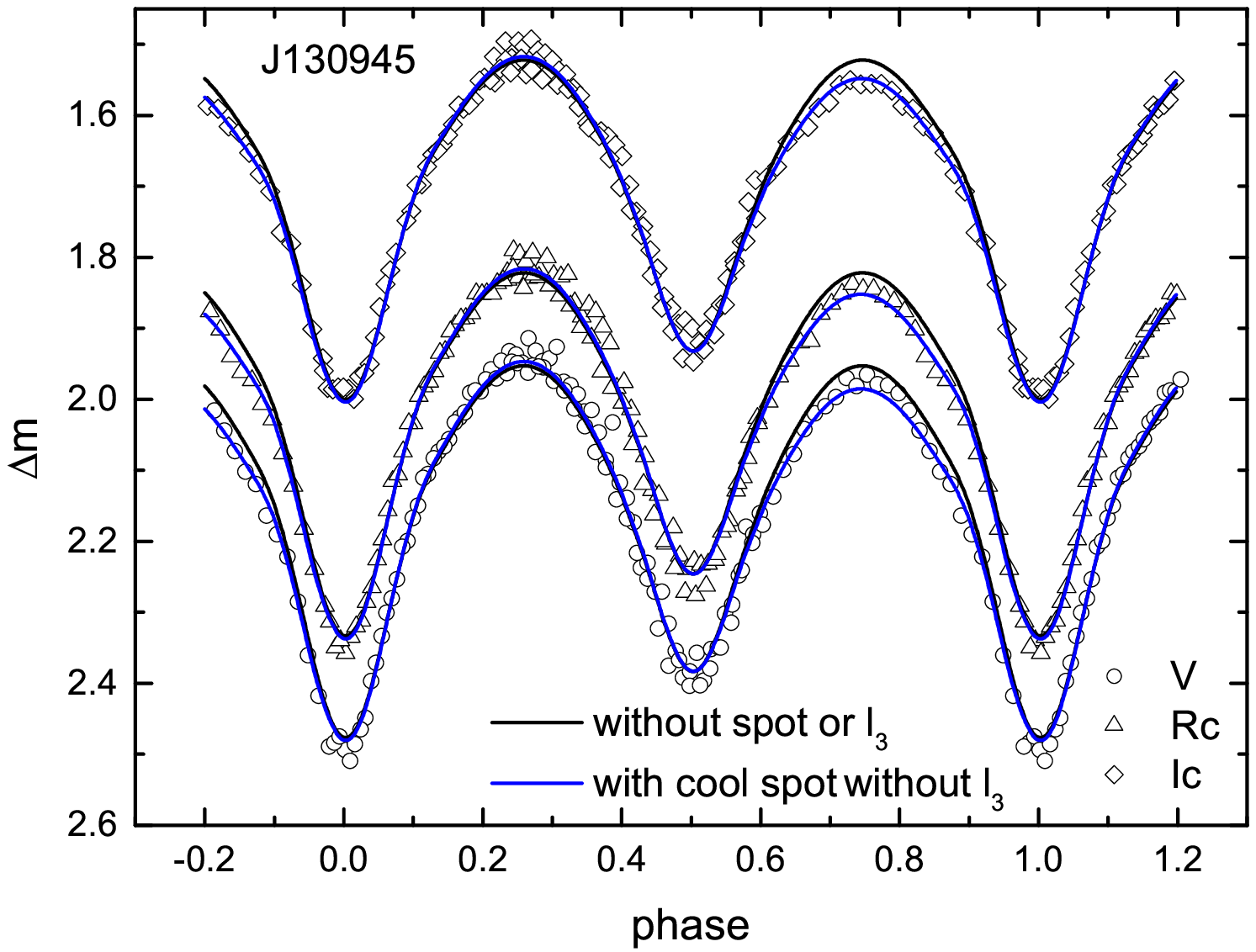}
\includegraphics[width=0.32\textwidth]{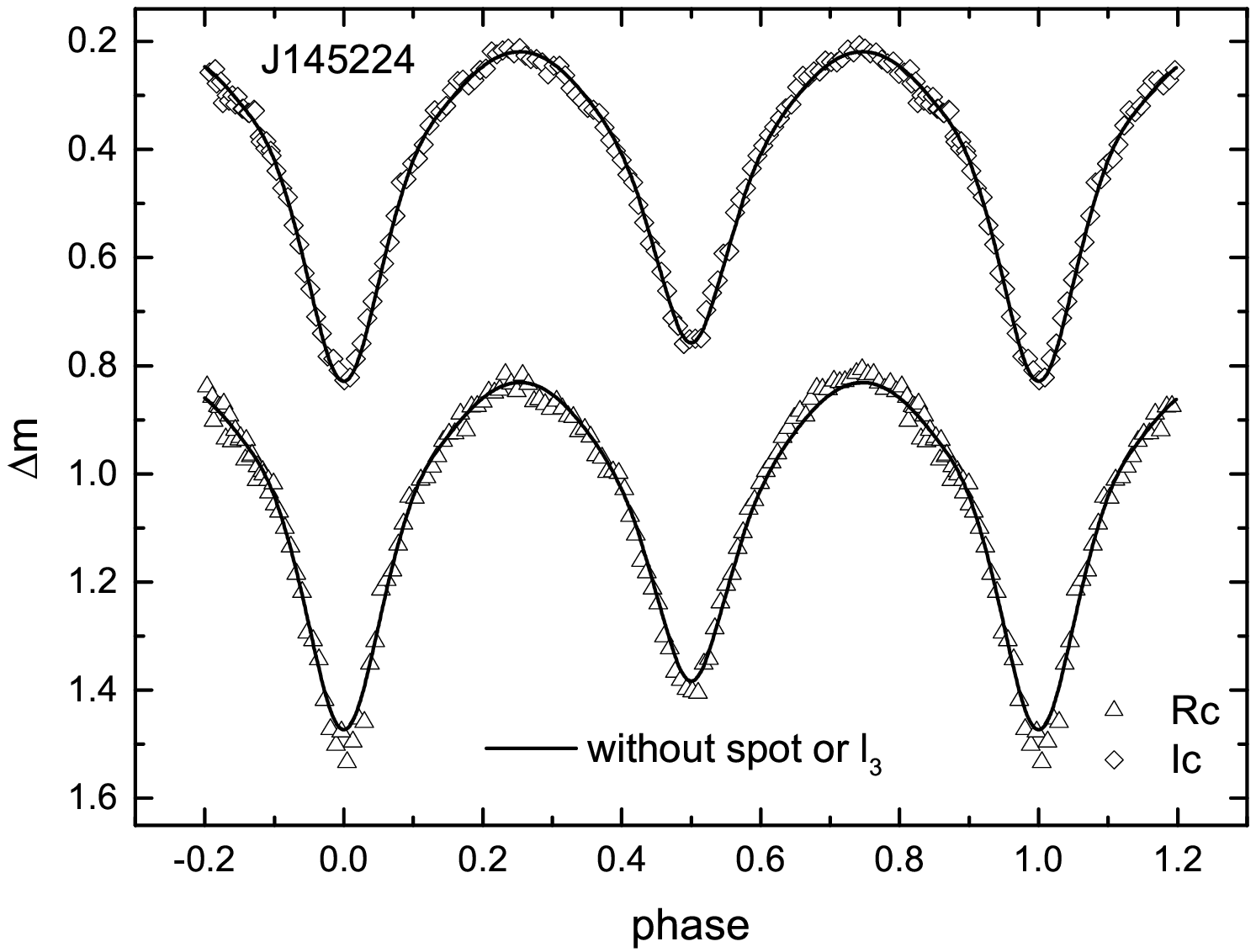}
\includegraphics[width=0.32\textwidth]{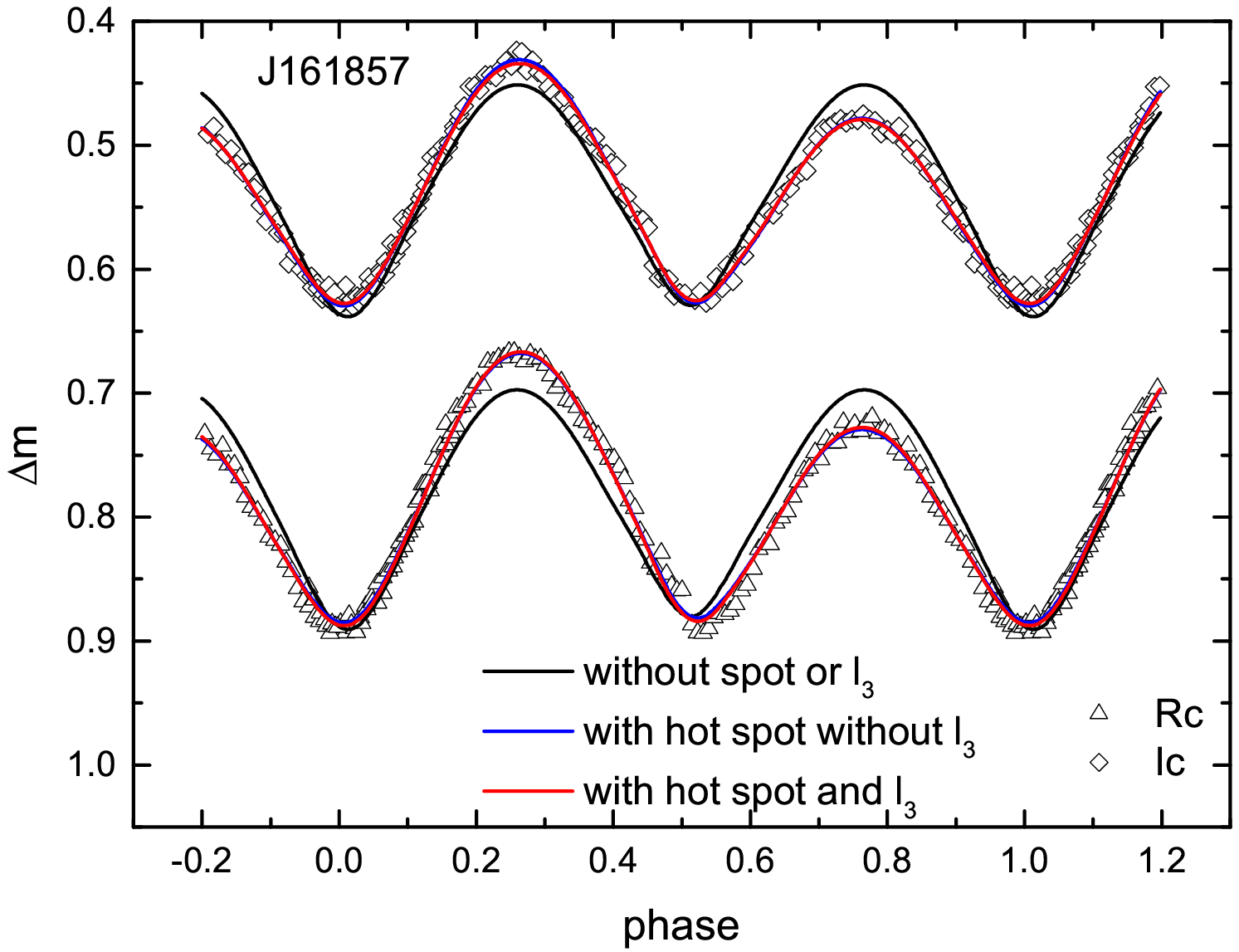}
\includegraphics[width=0.32\textwidth]{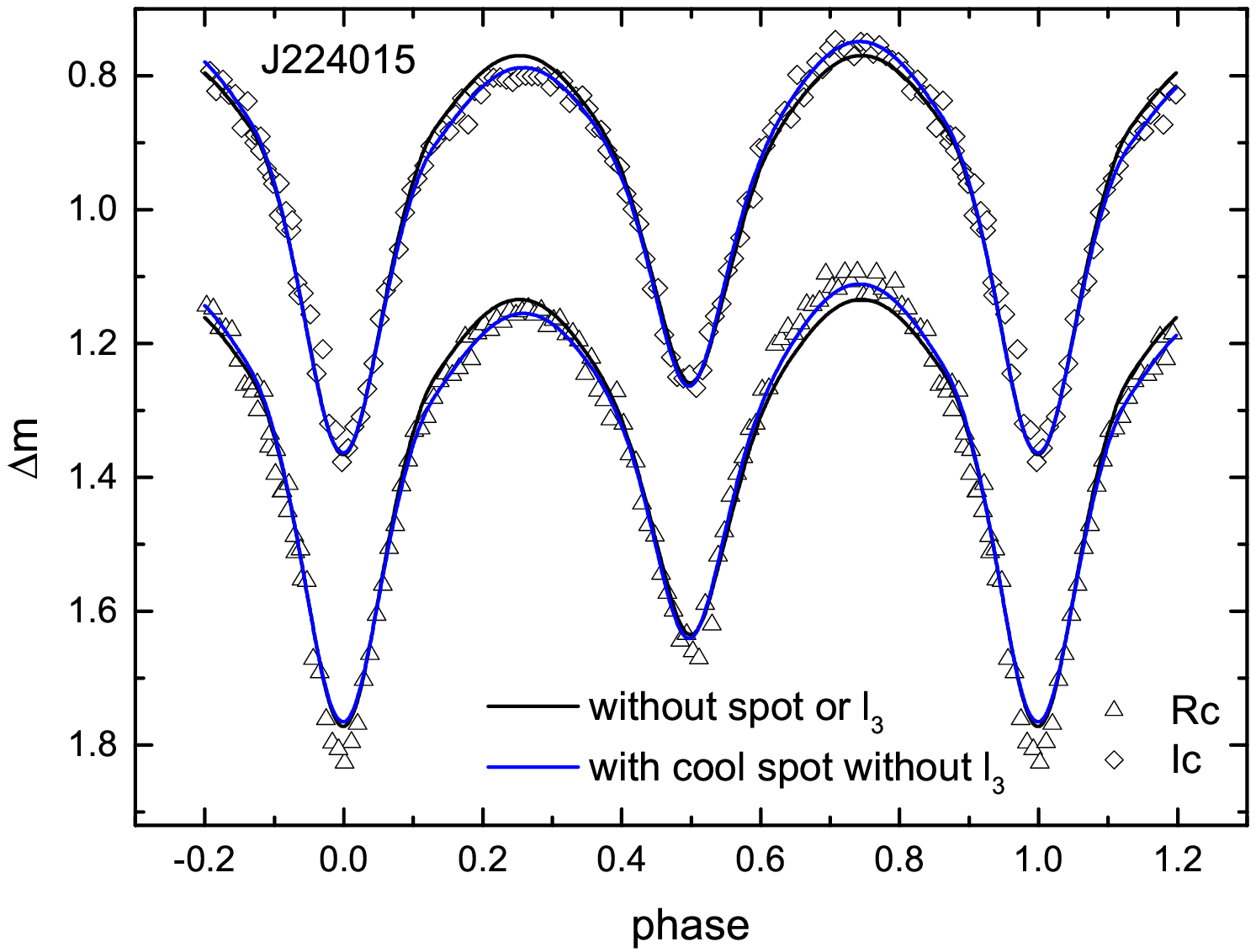}
\caption{Synthetic and observed light curves of the nine stars. The phases were calculated by using the period and HJD$_0$ presented in Table 1.
The symbols have the same meaning as Figure 1. The black, blue, and red lines represent the theoretical light curves with neither spot nor $l_3$, with a dark or hot spot but no $l_3$, and with $l_3$, respectively. }
\end{figure}

\renewcommand\arraystretch{1.3}
\begin{table*}
\tiny
\begin{center}
\caption{Photometric elemenets of the targets}
\begin{tabular}{p{0.8cm}|p{0.5cm}|p{1cm}|p{1cm}|p{0.9cm}|p{0.9cm}|p{1cm}|p{1cm}|p{1cm}|p{1cm}|p{1cm}|p{1cm}}
\hline
Star    &  Mode        &$q(M_2/M_1)$& $\Delta T(K)$  &  $i$      &   $\Omega$  &$l_{2V}/l_{1V}$&$l_{2R_c}/l_{1R_c}$ &$l_{2I_c}/l_{1I_c}$&$l_{3V}/l_{V}$ &$l_{3R_c}/l_{R_c}$&$l_{3I_c}/l_{I_c}$\\\hline
\multirow{2}{*}{J003033} &  1    & 2.064(92)  &    207(15)     & 82.3(5)   &   5.220(50) &   -        &  1.570(38) & 1.619(39)&    -         &    -       &    -         \\
        &  2  & 2.074(64)  &    179(12)      & 82.8(5)   &   5.215(40) &   -        &  1.615(27) & 1.657(27)&    -         &    -       &    -         \\
J014418 &  1        & 2.104(84)  &    212(13)     & 82.2(5)   &   5.308(55) &   -        &  1.523(31) & 1.592(33)&    -         &    -       &    -         \\
J031700 &  1        & 3.878(15)  &    177(6)      & 73.7(1)   &   7.730(20) &  2.688(20) &  2.838(17) & 2.929(16)&    -         &    -       &    -         \\
\multirow{2}{*}{J074350} &  1        & 2.297(99)  &    159(11)     & 80.0(5)   &   5.562(56) &   -        &  1.684(43) & 1.764(44)&    -         &    -       &    -         \\
        &  3  & 2.327(99)  &    159(12)     & 81.1(12)   &   5.601(55) &   -        &  1.703(48) & 1.784(52)&    -         & 0.9(2)\% &  2.9(2)\%  \\
\multirow{2}{*}{J104942} &  1        & 1.502(21)  &    103(23)     & 57.2(2)   &   4.486(33) &  1.231(22) &  1.273(19) & 1.305(15)&    -         &    -       &    -         \\
        &  3  & 1.449(23)  &    103(23)     & 59.6(6)   &   4.366(33) &  1.187(53) &  1.227(52) & 1.258(50)&  13.4(5)\%&15.9(5)\%& 17.5(5)\% \\
\multirow{2}{*}{J130945} &  1        & 3.662(109)   &    116(12)     & 72.4(4)   &   7.299(121) & 2.360(67)   &  2.408(66)  &2.550(62) &    -         &    -       &    -         \\
        &  2      & 3.657(82)   &    108(8)      & 73.1(3)   &   7.292(91) &  2.405(50)  &  2.440(49)  & 2.584(47) &    -         &    -       &    -         \\
J145224 &  1        & 1.935(38)  &    91(7)      & 77.0(2)   &   5.078(58) &   -        &  1.523(32) & 1.579(32)&    -         &    -       &    -         \\
\multirow{3}{*}{J161857} &  1        & 0.601(18)  &    74(89)     & 52.6(8)   &   3.042(33) &   -        &  0.571(29) & 0.582(25)&    -         &    -       &    -         \\
        &  2      & 0.602(7)   &    -179(35)    & 50.3(4)   &   3.055(14) &   -        &  0.777(16) & 0.748(13)&    -         &    -       &    -         \\
        &  4      & 0.597(8)&    -178(32)    & 50.8(5)   &   3.044(14) & -          &  0.771(28) & 0.742(27)&  -           & 0.6(3)\% &  4.4(3)\%  \\
\multirow{2}{*}{J224015} &  1        & 2.554(41)  &    236(19)     & 76.9(4)   &   5.975(49) &   -        &  1.725(30) & 1.836(27)&    -         &    -       &     -        \\
        &  2      & 2.501(97) &    202(13)     & 77.3(4)   &   5.860(70) &   -        &  1.760(45) & 1.854(47)&    -         &    -       &     -        \\\hline
Star    &  Mode        & $r_1$      &    $r_2$       & $f$ (\%)  &   Spot      &  $\theta$ (rad)  & $\lambda$ (rad)  & $r_s$ (rad)    & $T_s$        &$\Sigma$    &              \\\hline
\multirow{2}{*}{J003033} &  1        & 0.332(4)   &    0.456(19)   & 20.3(83)  &     -       &    -       &    -       &   -      &    -         & 0.0048     &              \\
        &  2      & 0.334(3)   &    0.448(14)   & 23.6(67)  &   Star 2    &  1.790(115)&  1.983(98)& 0.258(13)&  0.812(54)  & 0.0039     &              \\
J014418 &  1        & 0.325(3)   &    0.454(18)   & 15.1(92)  &    -        &    -       &    -       &   -      &    -         & 0.0075     &              \\
J031700 &  1        & 0.270(1)   &    0.502(2)    & 3.9(32)   &    -        &    -       &    -       &   -      &    -         & 0.0023     &              \\
\multirow{2}{*}{J074350} &  1        & 0.320(3)   &    0.459(23)   & 17.3(98) &    -        &    -       &    -       &   -      &    -         & 0.0020     &              \\
        &  3  & 0.319(4)   &    0.459(19)   & 17.7(91)  &    -        &    -       &    -       &   -      &    -         & 0.0017     &              \\
\multirow{2}{*}{J104942} &  1        & 0.349(2)   &    0.418(6)    & 7.4(56)   &    -        &    -       &    -       &   -      &    -         & 0.0047     &              \\
        &  3  & 0.357(2)   &    0.421(7)    & 14.6(57)  &    -        &    -       &    -       &   -      &    -         & 0.0042     &              \\
\multirow{2}{*}{J130945} &  1        & 0.288(2)   &    0.504(13)   & 28.8(193)  &    -        &    -       &    -       &   -      &    -         & 0.0010     &              \\
        &  2      & 0.288(1)   &    0.506(10)    & 29.0(145)  &   Star 1    &  1.788(110)&  1.456(88)& 0.375(23)&  0.748(45)  & 0.0008     &              \\
J145224 &  1        & 0.330(3)   &    0.451(13)   & 13.7(98) &    -        &    -       &    -       &   -      &    -         & 0.0023     &              \\
\multirow{2}{*}{J161857} &  1        & 0.428(3)   &    0.338(14)   & 6.7(94)   &    -        &    -       &    -       &   -      &    -         & 0.0035     &              \\
        &  2      & 0.425(1)   &    0.339(6)    & 3.5(40)   &   Star 2    &  1.643(95) &  1.287(112) & 0.349(59)&  1.260(98)  & 0.0012     &              \\
        &  4& 0.427(2)  &    0.334(6)    & 3.8(41)   &   Star 2    &  1.643     &  1.287     & 0.349    &  1.260       & 0.0010     &              \\
\multirow{2}{*}{J224015} &  1        & 0.304(2)   &    0.468(6)   & 7.0(81)  &    -        &    -       &    -       &   -      &    -         & 0.0029     &              \\
        &  2      & 0.308(4)   &    0.474(18)   & 14.2(115) &   Star 2    &  1.607(145)&  1.899(88)& 0.283(19)&  0.814(65)   & 0.0021     &              \\\hline
\end{tabular}
\end{center}
Note$-$ 'Mode 1' means solutions with neither spot nor l$_3$, 'Mode 2' represents solutions with spot but no l$_3$, 'Mode 3' refers to solutions with l$_3$ but no spot, 'Mode 4' represents solutions with spot and l$_3$.
\end{table*}

\renewcommand\arraystretch{1.3}
\begin{table*}
\centering
\scriptsize
\caption{The final temperatures of the two components of all the targets}
\begin{tabular}{lcccccccccccc}
\hline
Star     & J003033 &J014418   & J031700  &J074350  & J104922  &J130945  &J145224   & J161857 & J224015   \\\hline
$T_1$(K)  &5246(17)& 4819(282)&4968(206) &4471(552)&4560(566) &3762(268)&3898(106) &4400(610)&4643(284)\\
$T_2$(K)  &5067(29)& 4607(295)&4791(212) &4312(564)&4457(589) &3654(276)&3807(113) &4578(642)&4441(297) \\
\hline
\end{tabular}
\end{table*}

\section{Orbital Period variations study}
The period variation study of eclipsing binaries is a powerful tool to investigate their dynamical evolutions and to search for additional companions (e.g., Zhou et al. 2016b; Er-gang et al. 2019; Liao et al. 2019). To analyze the orbital period variations of the nine targets, we tried to collect as many eclipsing times as possible. Some worldwide photometric surveys, such as Northern Sky Variability Survey (NSVS, Wozniak 2004), Wide Angle Search for Planets (SuperWASP, Butters et al. 2010), LINEAR (Palaversa et al. 2013), Catalina Sky Surveys (CSS, Drake et al. 2014b), and All-Sky Automated Survey for SuperNovae (ASAS-SN, Shappee et al. 2014; Jayasinghe et al. 2018) have already observed most of our targets. However, the time resolution of NSVS, LINEAR, CSS, and ASAS-SN is very low. We constructed the complete phase or half of phase light curve using longer time span data. Using these light curves, the eclipsing times were calculated with the K-W method (Kwee \& van Woerden 1956) and are displayed in Table 6. Figure 4 displays an example for deriving the complete phase or half of phase light curve. First, we divided the CSS light curves of J073450 into four parts as shown in the upper panel of Figure 4. Second, using the equation, $MJD=MJD_0+P\times E$, where $MJD$ is the observing time, $MJD_0$ is the reference time for each part, and $P$ is the orbital period of J073450, we moved the long time span data into one period and determined the four lower panels of Figure 4. Then, we obtained seven eclipsing times from these four sets of light curves. Since the observing times of NSVS, LINEAR and CSS are given in $MJD$, we firstly transformed $MJD$ to $JD$ using $JD = MJD+2400000.5$, and then converted to $HJD$.
Four targets (J003033, J031700, J104942, and J161857) have been observed by SuperWASP, and Lohr et al.(2013b) have obtained their eclipsing times. We acquired the eclipsing times of the four targets thank to a private communication with Dr. Lohr. Using the following equation,
\begin{equation}
HJD=HJD_0+P\times E,
\end{equation}
we computed the $O-C$ values for all the systems. In this equation, $HJD$ is the time of minimum light which is listed in Table 6, $HJD_0$ is the initial epoch which is shown in Table 1, and $P$ is the orbital period which is also shown in Table 1. The $O-C$ diagrams of all targets are displayed in Figure 5. For J031700, we found an obvious linear variation from cycle -12372 to -5097 (eight points) which is shown in a dashed line square box. This means that the period of J031700 is incorrect. A linear equation was applied to fit the eight points and a new period was determined to be 0.2256396(3) days. Then, we used this orbital period to reconstruct the $O-C$ diagram of J031700 which is also plotted in Figure 5. Now, we can see that all the $O-C$ diagrams manifest long-term orbital period changes. The following second-order polynomial,
\begin{eqnarray}
O-C= \Delta T_{0} + \Delta P_{0}\times E+{\beta \over 2}E^2,
\end{eqnarray}
was applied to fit the $O-C$ diagrams. The results are shown in Table 7. We have found that five systems exhibit long-term increasing orbital period and the others manifest secular decreasing orbital period.

\begin{figure*}
\begin{center}
\includegraphics[width=0.6\textwidth]{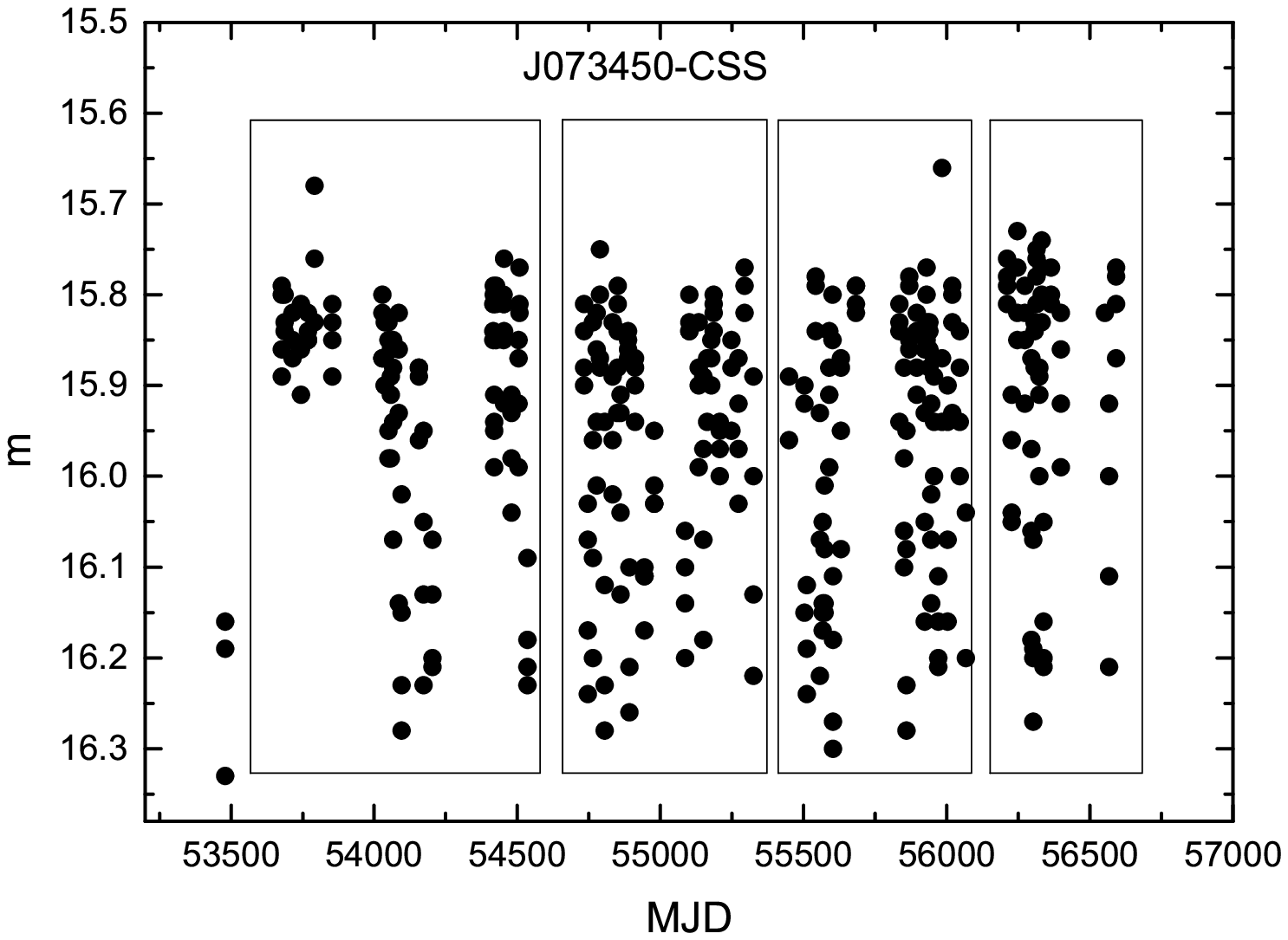}\\
\includegraphics[width=0.235\textwidth]{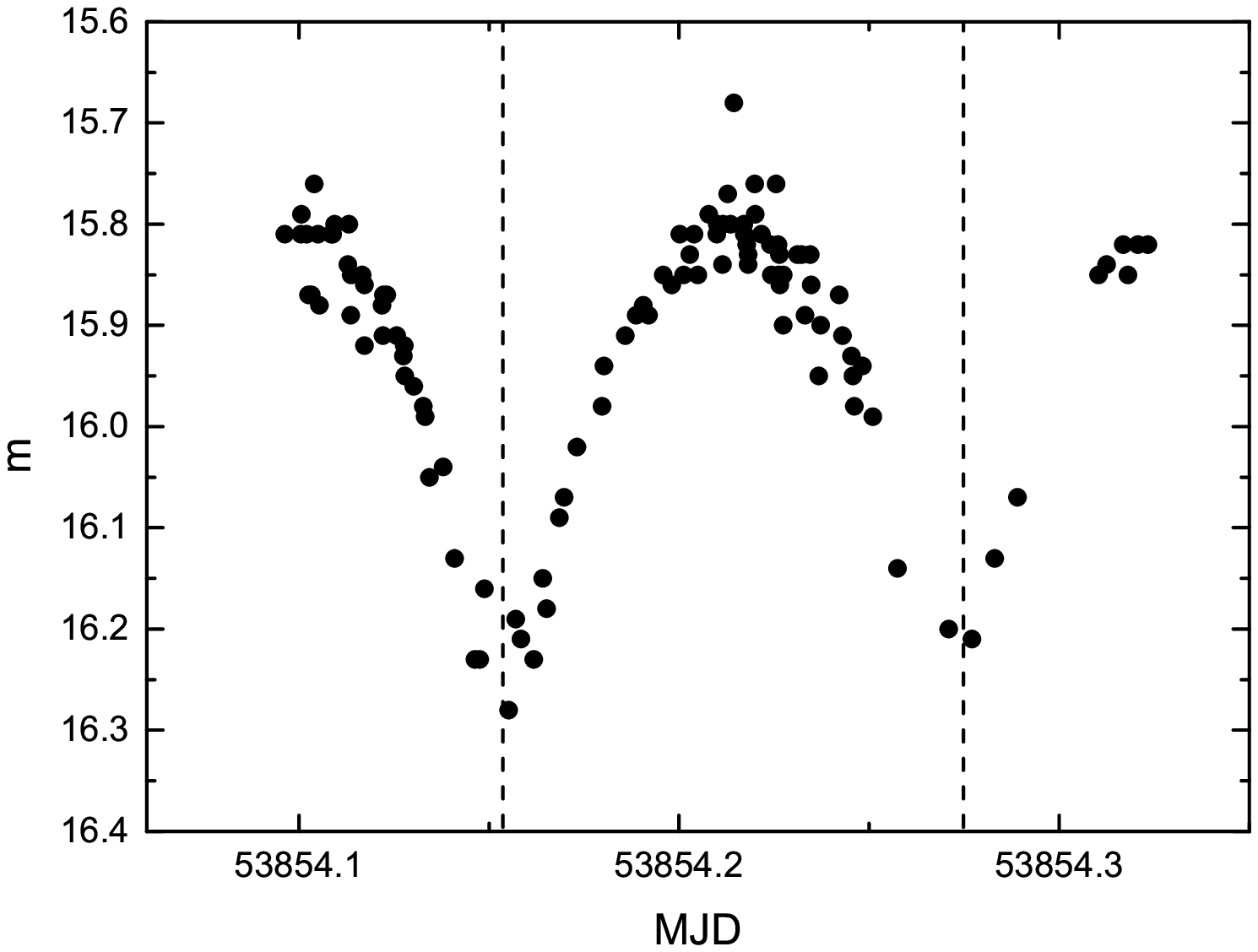}
\includegraphics[width=0.25\textwidth]{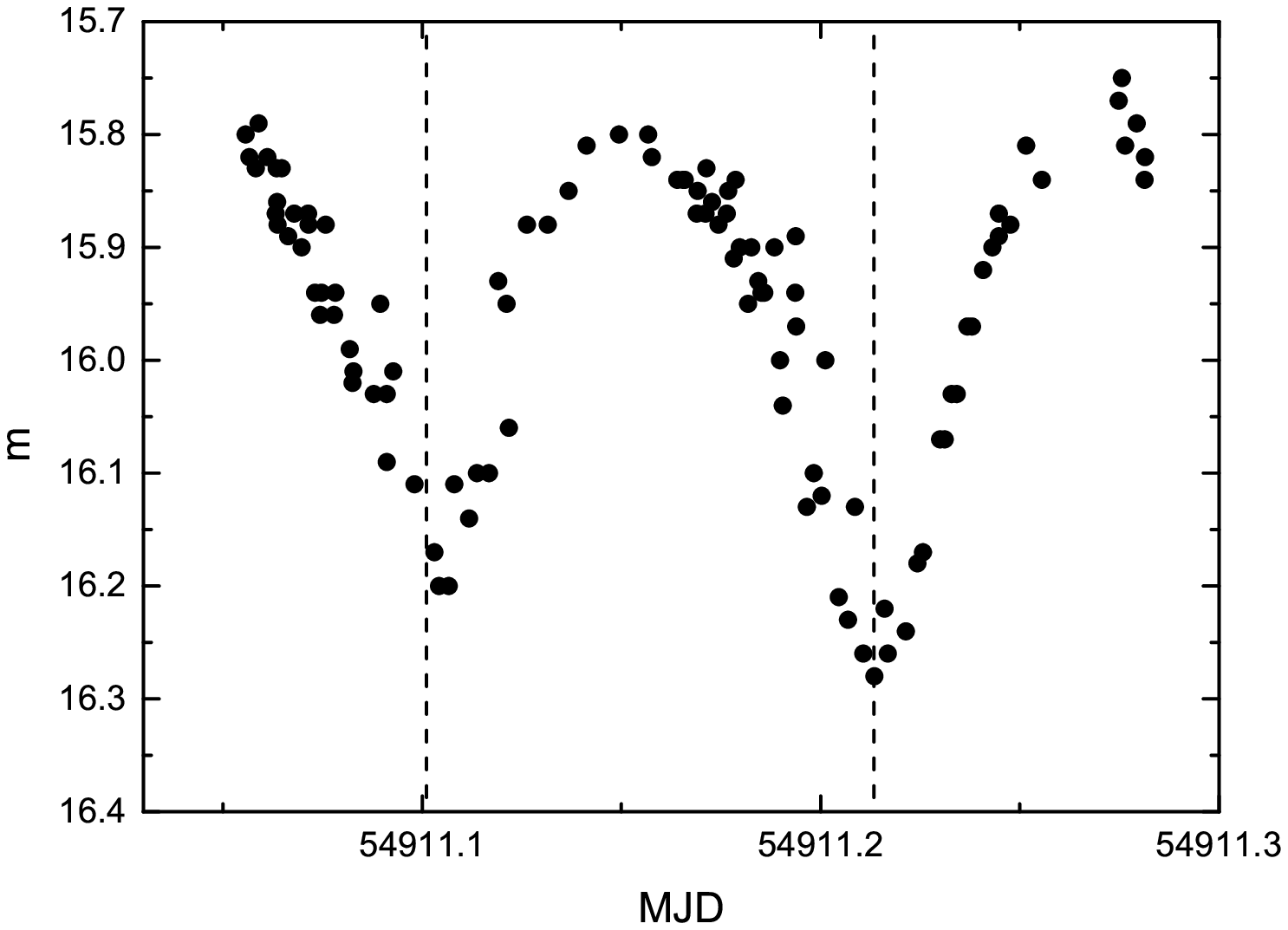}
\includegraphics[width=0.235\textwidth]{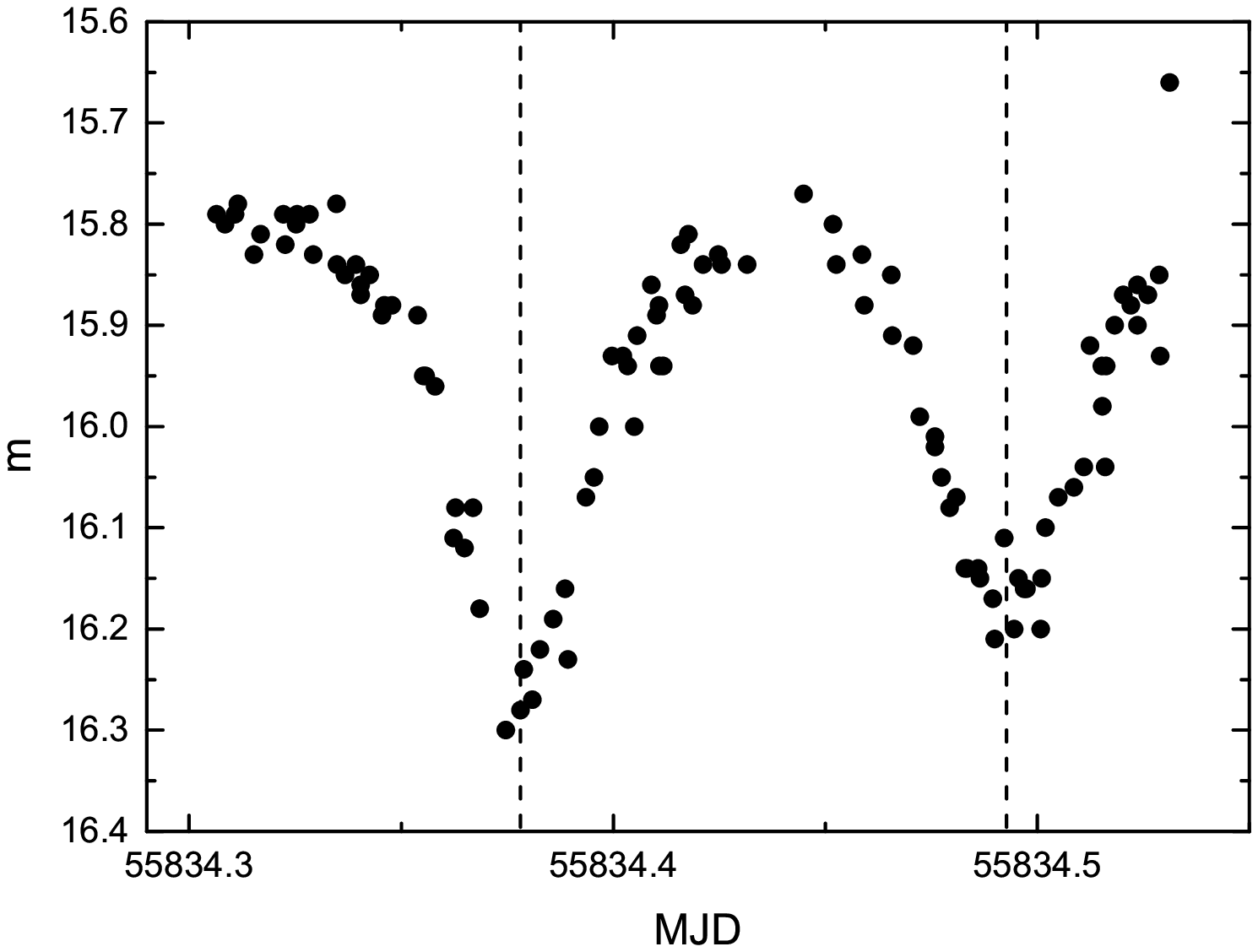}
\includegraphics[width=0.235\textwidth]{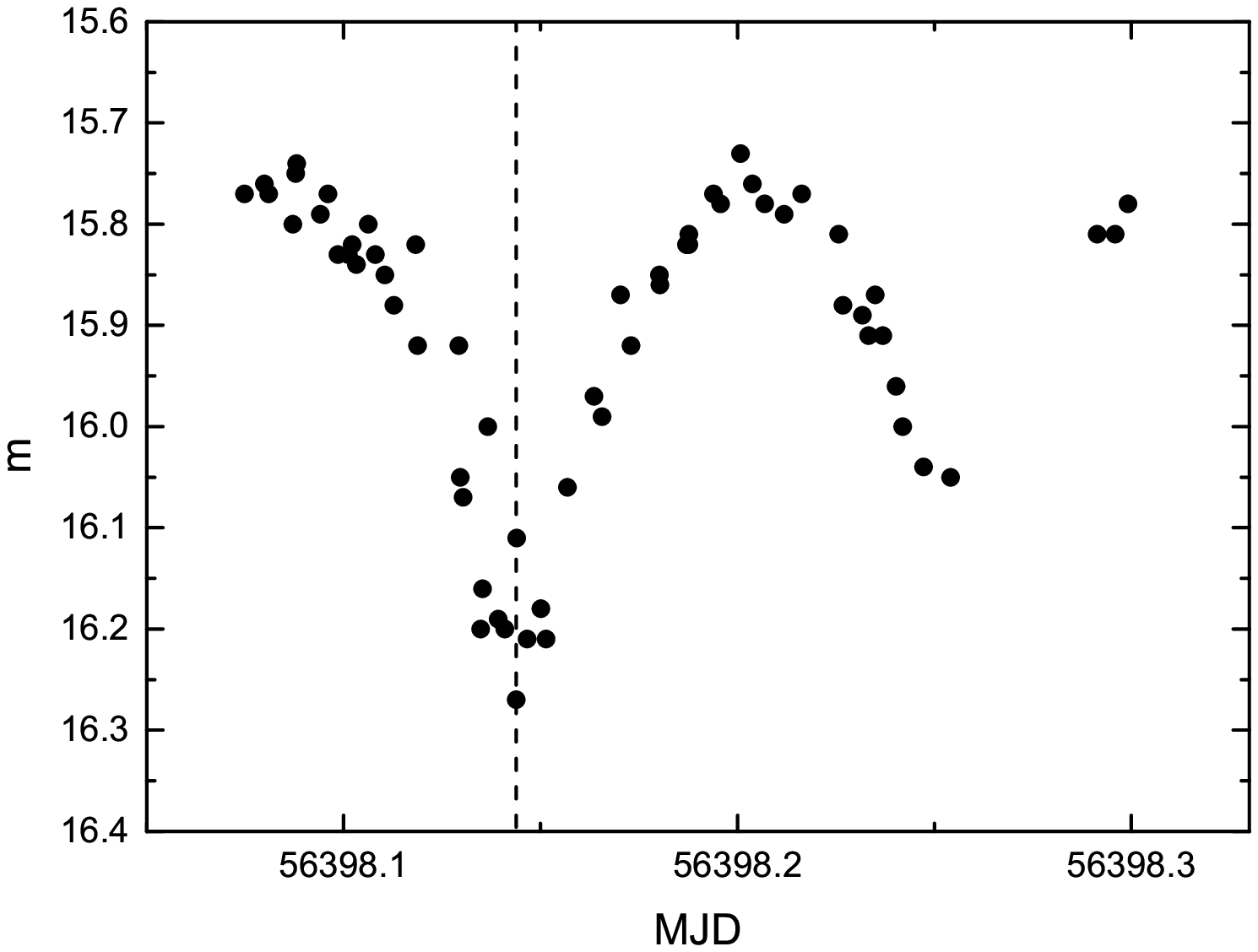}

\caption{The upper panel plots the CSS light curves of J073450, we divided the light curves into four parts. Using the equation, $MJD=MJD_0+P\times E$, where $MJD$ is the observing time, $MJD_0$ is the reference time for each part, and $P$ is the orbital period, we moved the long time span data into one period and determined the four lower panels.}
\end{center}
\end{figure*}

\begin{figure}
\includegraphics[width=0.32\textwidth]{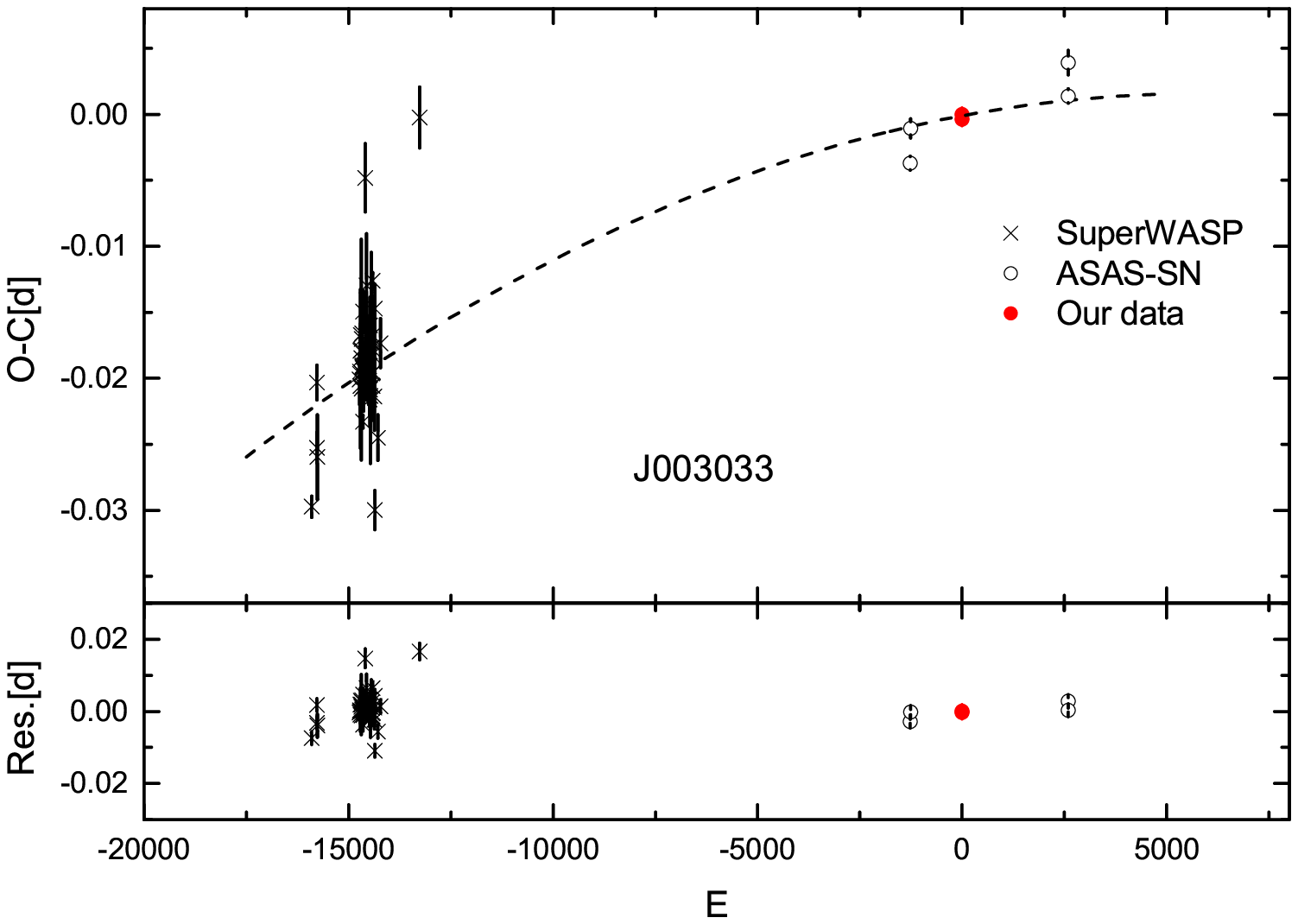}
\includegraphics[width=0.32\textwidth]{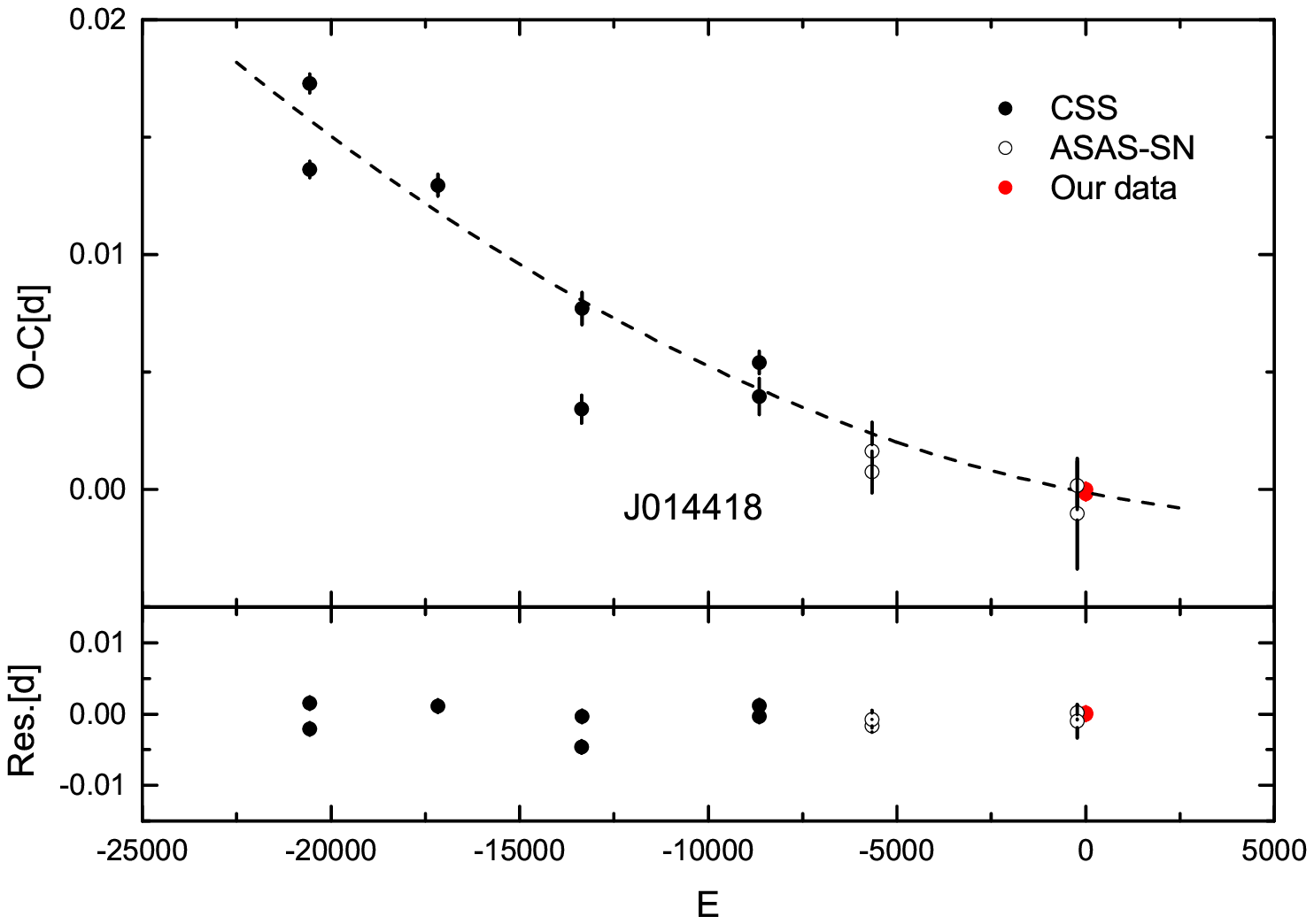}
\includegraphics[width=0.32\textwidth]{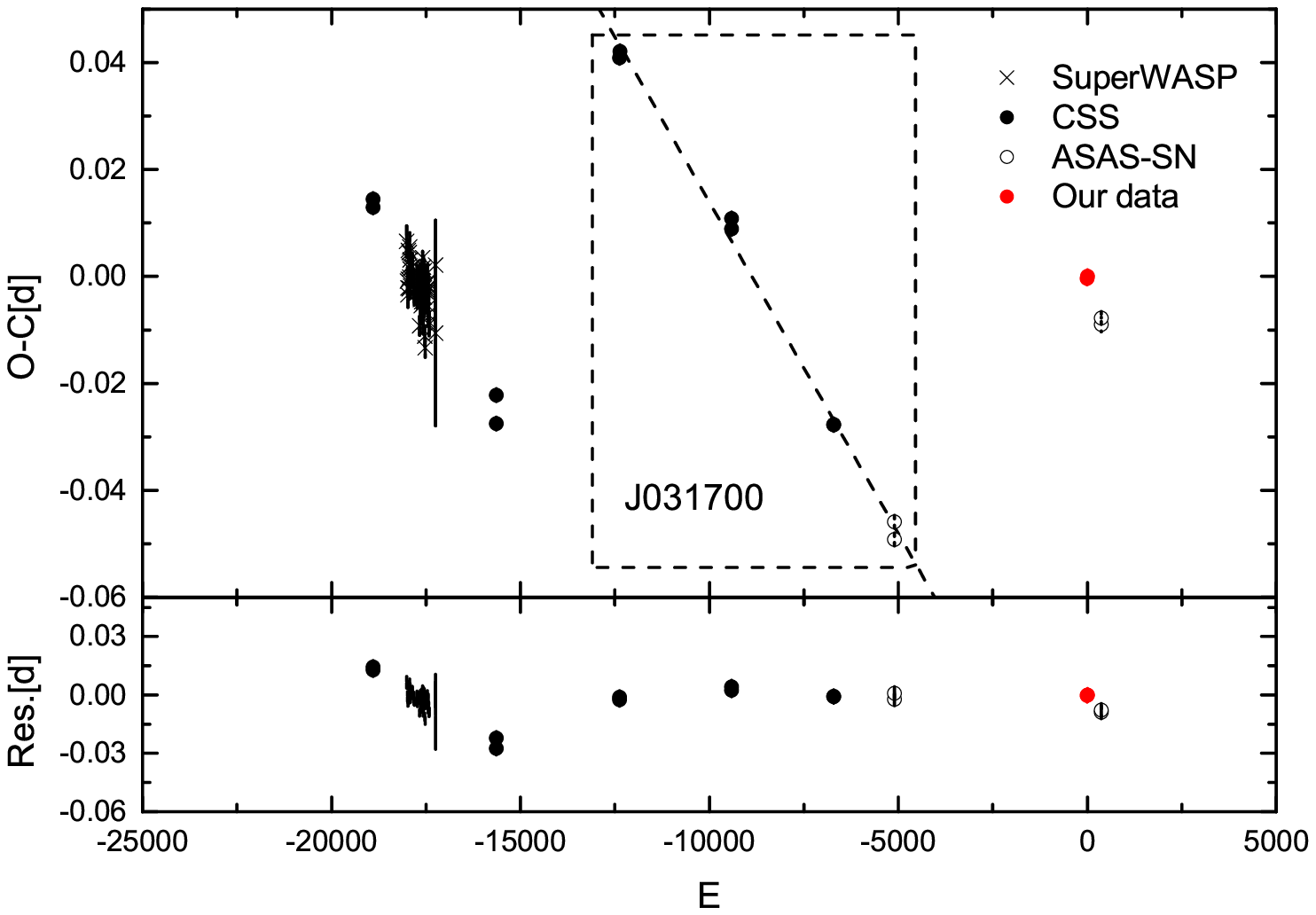}
\includegraphics[width=0.32\textwidth]{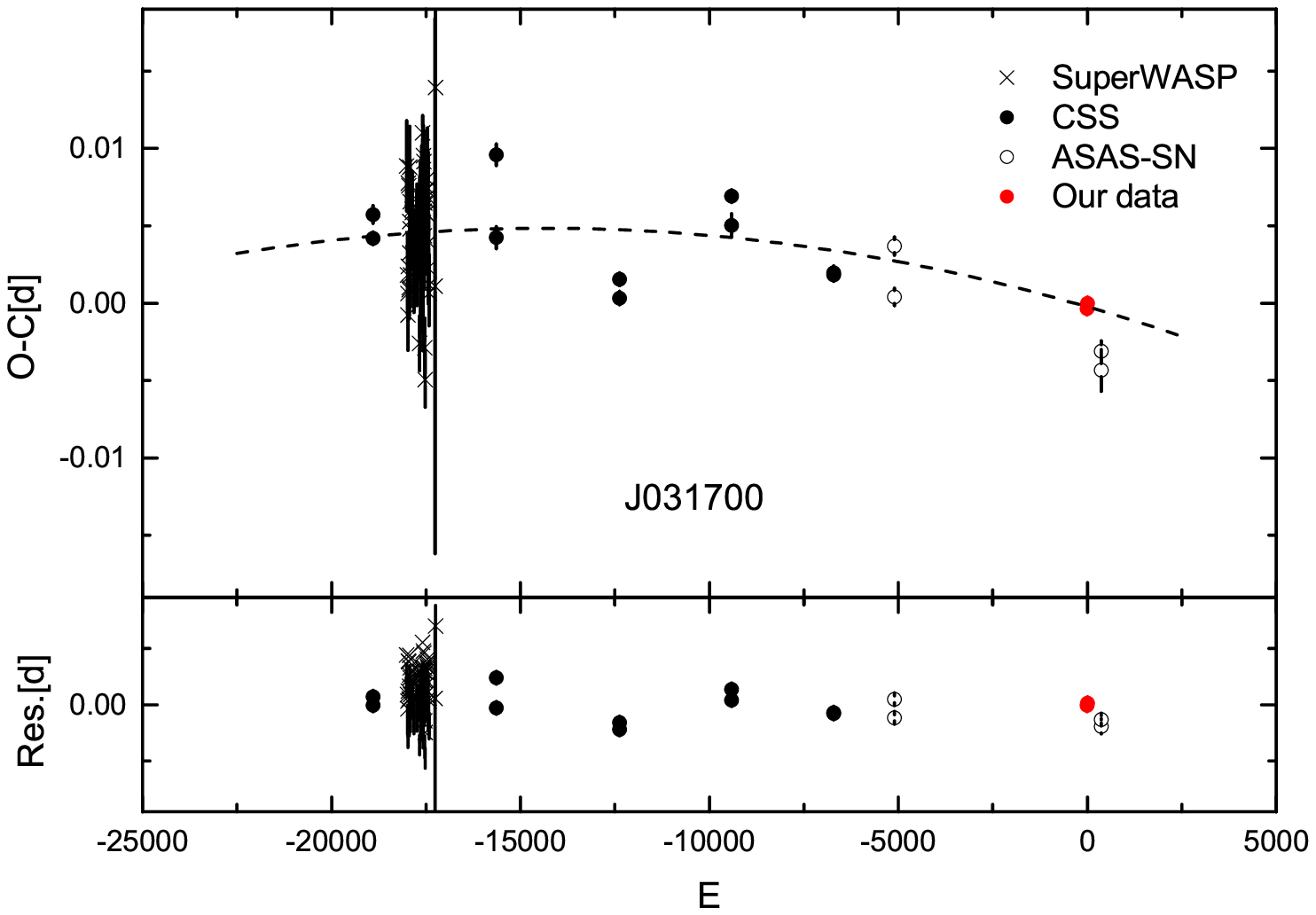}
\includegraphics[width=0.32\textwidth]{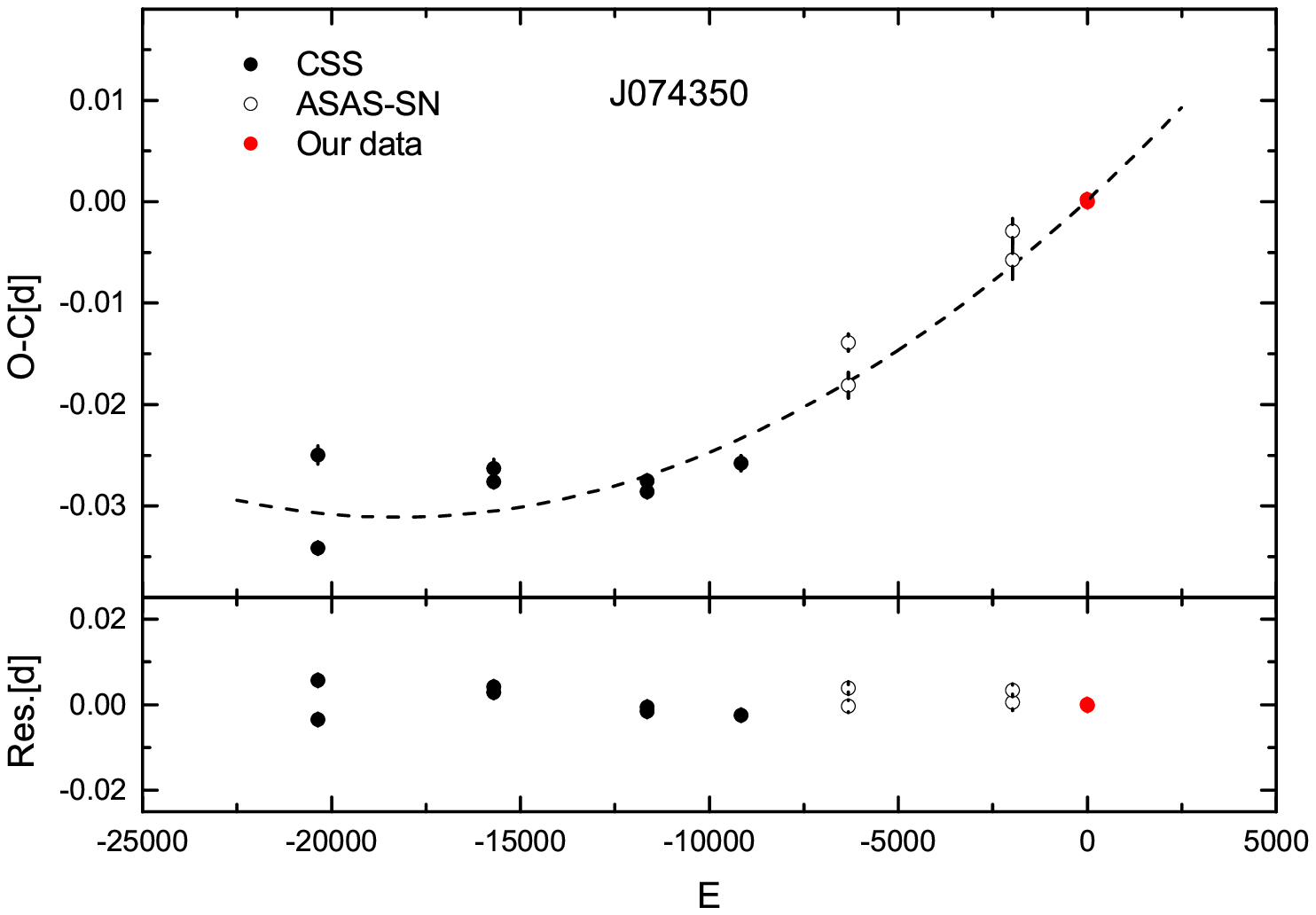}
\includegraphics[width=0.32\textwidth]{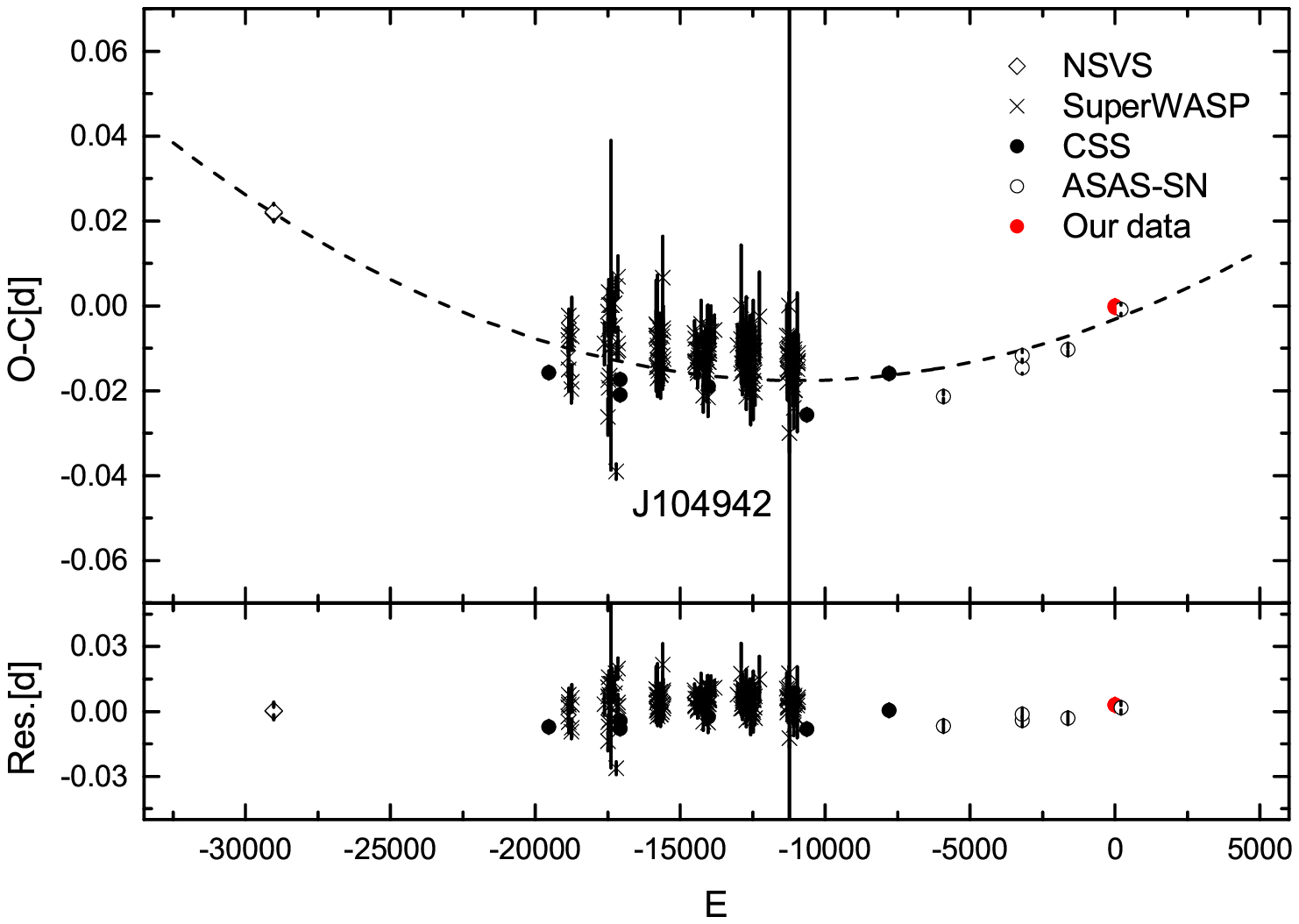}
\includegraphics[width=0.32\textwidth]{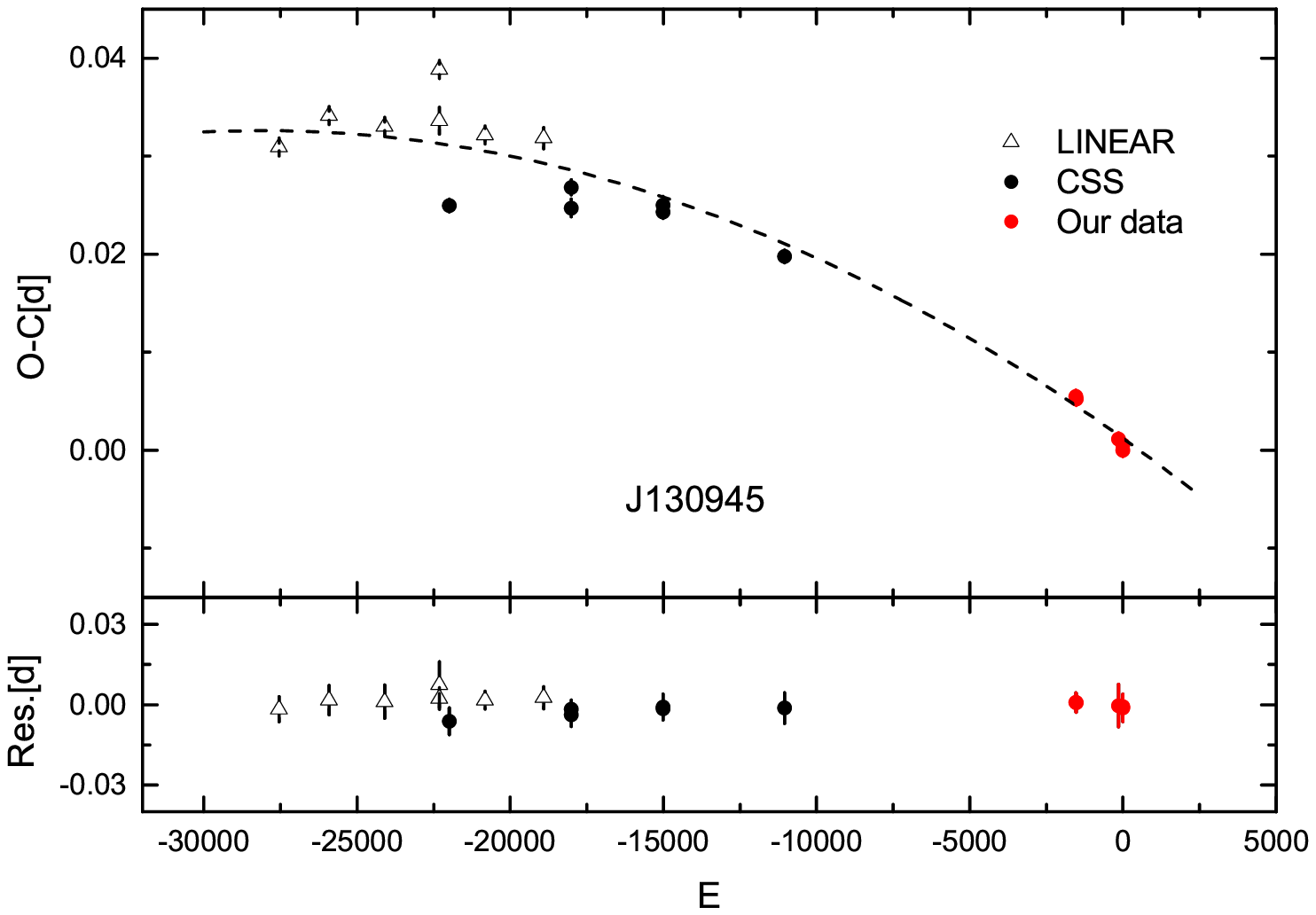}
\includegraphics[width=0.32\textwidth]{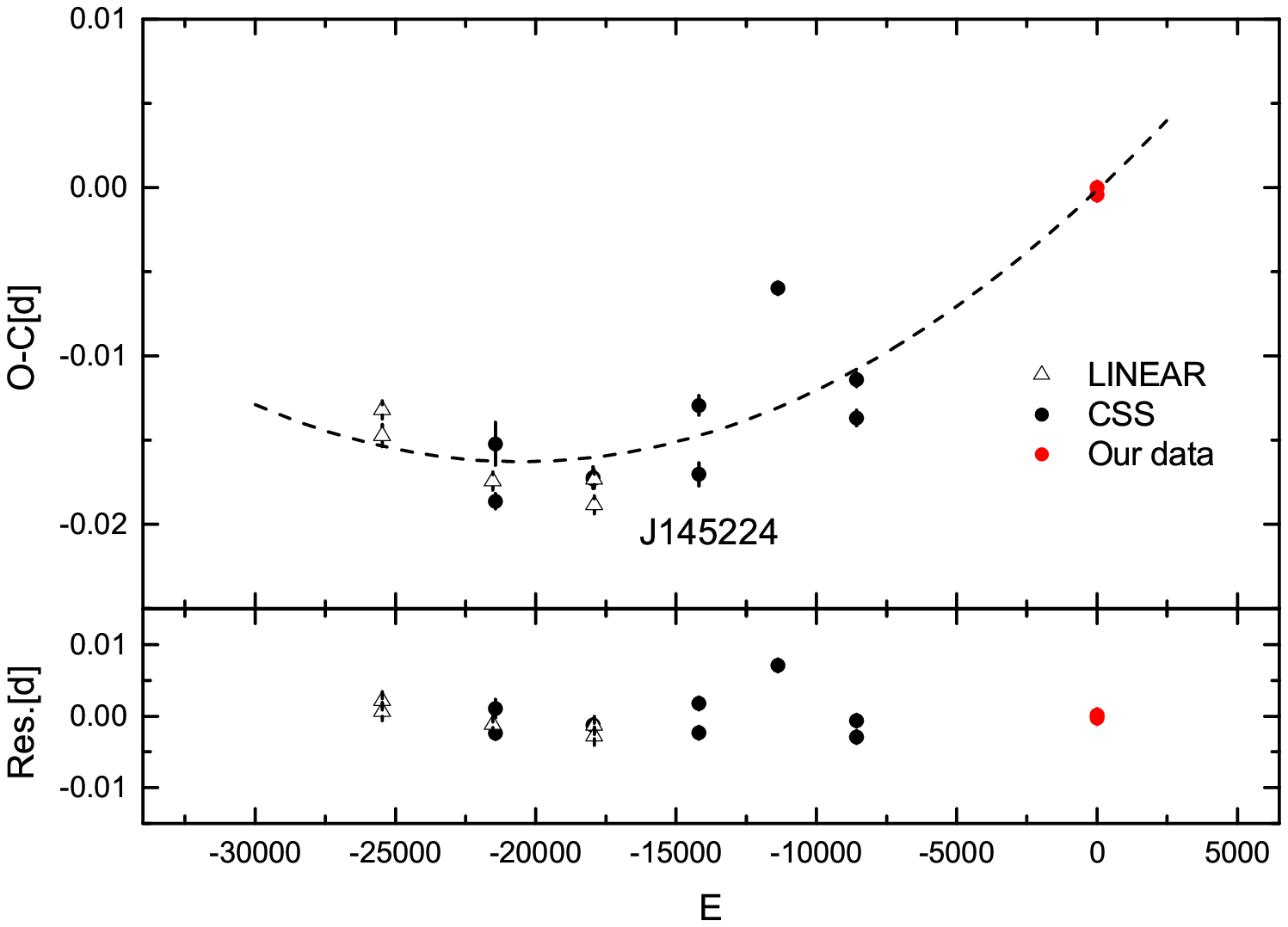}
\includegraphics[width=0.32\textwidth]{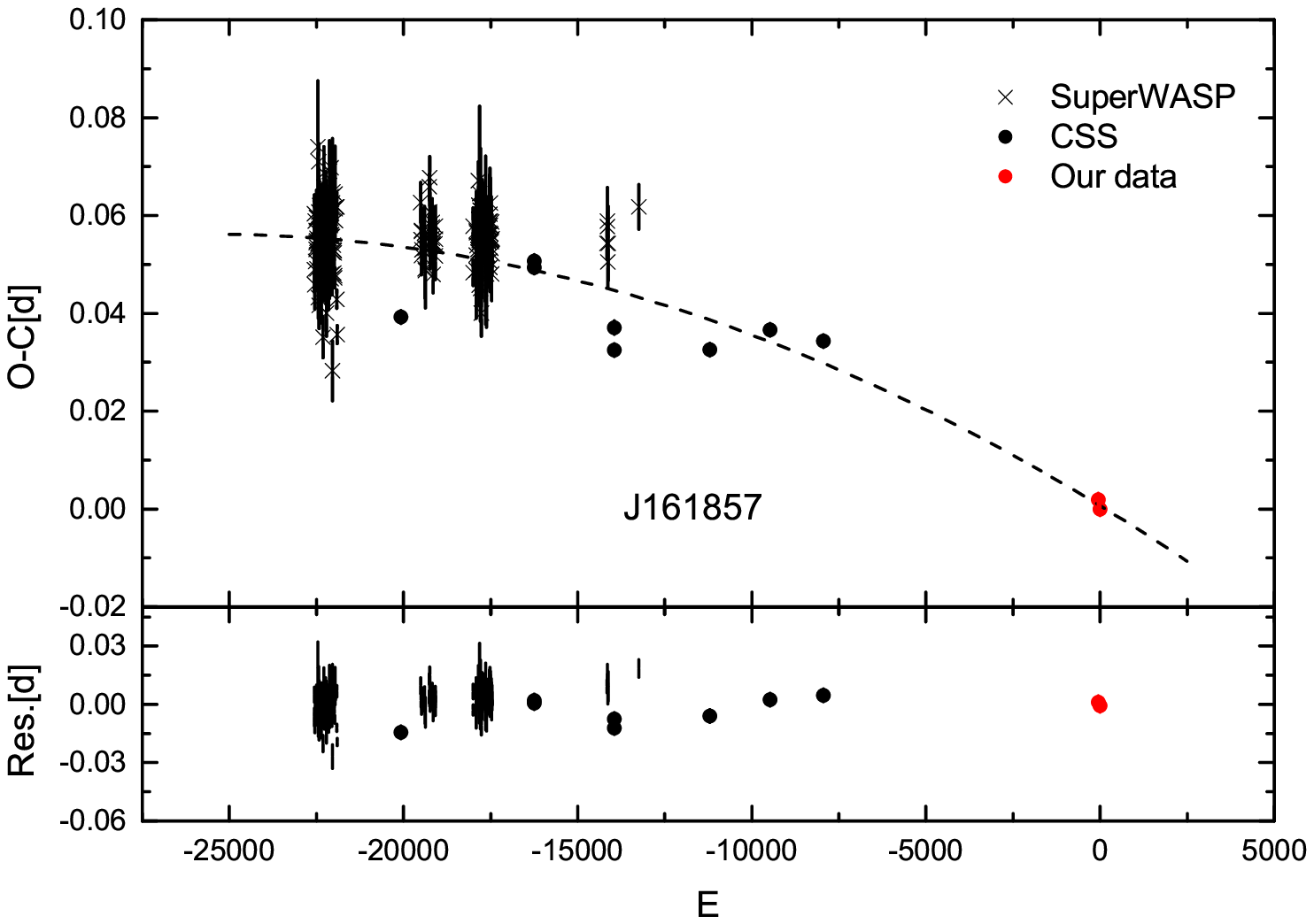}
\includegraphics[width=0.32\textwidth]{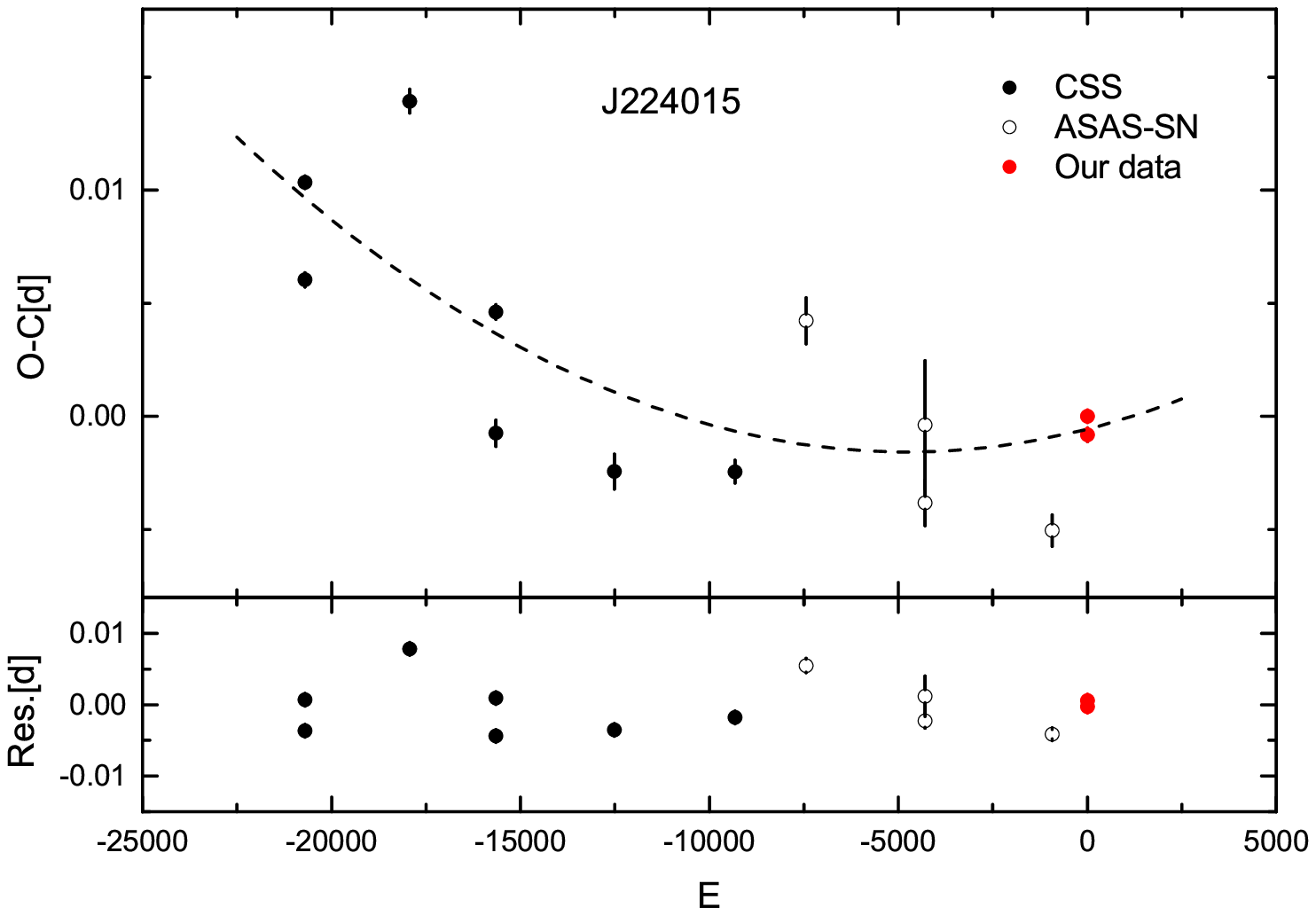}
\caption{The $O-C$ diagrams of the nine eclipsing binaries studied. }
\end{figure}

\renewcommand\arraystretch{1.2}
\begin{table*}
\tiny
\caption{The eclipse timings for the nine stars}
\begin{tabular}{p{0.6cm}p{1cm}p{0.9cm}p{1cm}p{1cm}p{1cm}p{0.3cm}|p{0.6cm}p{1cm}p{0.9cm}p{1cm}p{1cm}p{1cm}p{0.3cm}}
\hline
Star     &  HJD        &  Error    &  E        & O-C      &  Residual &  Ref.& Star     &  HJD        &  Error    &  E        & O-C      &  Residual &  Ref. \\ \hline
         &  2450000+   &           &           &          &           &      &          &  2450000+   &           &           &          &           &        \\ \hline
J003033  &  7388.66706 &  0.00052  &  -1264.5  & -0.00371 &  -0.00279 &  (1) & J130945  &  2724.55607 &  0.00018  &  -23683   & -0.00968 &  0.00320  &  (5) \\
         &  7388.78301 &  0.00072  &  -1264    & -0.00107 &  -0.00015 &  (1) &          &  3070.71019 &  0.00015  &  -23682.5 & -0.01442 &  -0.00154 &  (5) \\
         &  7675.22923 &  0.00015  &  0        & 0.00000  &  0.00010  &  (2) &          &  3453.59701 &  0.00092  &  -20380   & -0.01069 &  0.00155  &  (5) \\
         &  7675.11556 &  0.00037  &  -0.5     & -0.00036 &  -0.00026 &  (2) &          &  3829.73445 &  0.00051  &  -20379.5 & -0.00968 &  0.00256  &  (5) \\
         &  8262.06045 &  0.00093  &  2589.5   & 0.00391  &  0.00285  &  (1) &          &  3829.83478 &  0.00044  &  -19068.5 & -0.00706 &  0.00481  &  (5) \\
         &  8262.17124 &  0.00054  &  2590     & 0.00139  &  0.00033  &  (1) &          &  3899.71082 &  0.00022  &  -17203   & -0.00944 &  0.00181  &  (3) \\
J014418  &  3643.81909 &  0.00035  &  -20560   & 0.01363  &  -0.00209 &  (3) &          &  4144.73673 &  0.00090  &  -17202.5 & -0.01096 &  0.00029  &  (5) \\
         &  3643.93143 &  0.00040  &  -20559.5 & 0.01728  &  0.00156  &  (3) &          &  4548.63191 &  0.00040  &  -15267   & -0.01012 &  0.00036  &  (5) \\
         &  4381.79635 &  0.00030  &  -17165   & 0.01295  &  0.00113  &  (3) &          &  4739.69924 &  0.00073  &  -12077   & -0.01256 &  -0.00365 &  (3) \\
         &  4381.90504 &  0.00047  &  -17164.5 & 0.01295  &  0.00114  &  (3) &          &  4739.80691 &  0.00041  &  -8879.5  & -0.00900 &  -0.00202 &  (3) \\
         &  5210.51757 &  0.00060  &  -13352.5 & 0.00342  &  -0.00462 &  (3) &          &  5370.56193 &  0.00114  &  -6582    & -0.00253 &  0.00284  &  (3) \\
         &  5210.63054 &  0.00069  &  -13352   & 0.00770  &  -0.00033 &  (3) &          &  5370.66680 &  0.00066  &  -4978    & -0.00069 &  0.00344  &  (3) \\
         &  6233.79679 &  0.00077  &  -8645    & 0.00395  &  -0.00031 &  (3) &          &  6209.80643 &  0.00118  &  -4977.5  & -0.00058 &  0.00356  &  (3) \\
         &  6233.90693 &  0.00047  &  -8644.5  & 0.00540  &  0.00115  &  (3) &          &  8214.91272 &  0.00103  &  -3387.5  & -0.00300 &  -0.00018 &  (2) \\
         &  6881.88820 &  0.00089  &  -5663.5  & 0.00074  &  -0.00163 &  (1) &          &  8217.65720 &  0.00093  &  -3387    & 0.00123  &  0.00406  &  (2) \\
         &  6881.99778 &  0.00122  &  -5663    & 0.00164  &  -0.00073 &  (1) &          &  8510.91545 &  0.00136  &  -1753.5  & -0.00243 &  -0.00105 &  (2) \\
         &  8064.93473 &  0.00103  &  -221     & 0.00016  &  0.00022  &  (1) &          &  8540.89560 &  0.00059  &  -1753    & -0.00461 &  -0.00324 &  (2) \\
         &  8065.04222 &  0.00235  &  -220.5   & -0.00103 &  -0.00097 &  (1) &          &  8541.00062 &  0.00166  &  -176.5   & 0.00145  &  0.00135  &  (2) \\
         &  8112.97378 &  0.00021  &  0        & 0.00000  &  0.00013  &  (2) & J145224  &  2894.64409 &  0.00065  &  -25469.5 & -0.01474 &  0.00063  &  (5) \\
         &  8113.08228 &  0.00016  &  0.5      & -0.00019 &  -0.00005 &  (2) &          &  2894.74363 &  0.00030  &  -25469   & -0.01320 &  0.00216  &  (5) \\
J031700  &  3786.60872 &  0.00057  &  -18899.5 & 0.00573  &  0.00140  &  (3) &          &  3665.76047 &  0.00039  &  -21535.5 & -0.01743 &  -0.00119 &  (5) \\
         &  3786.72000 &  0.00030  &  -18899   & 0.00419  &  -0.00014 &  (3) &          &  3685.65469 &  0.00045  &  -21434   & -0.01863 &  -0.00238 &  (3) \\
         &  4523.54616 &  0.00071  &  -15633.5 & 0.00424  &  -0.00056 &  (3) &          &  3685.75611 &  0.00127  &  -21433.5 & -0.01522 &  0.00103  &  (3) \\
         &  4523.66434 &  0.00069  &  -15633   & 0.00960  &  0.00480  &  (3) &          &  4365.62865 &  0.00064  &  -17965   & -0.01724 &  -0.00122 &  (3) \\
         &  5259.57863 &  0.00043  &  -12372   & 0.00033  &  -0.00441 &  (3) &          &  4365.72660 &  0.00042  &  -17964.5 & -0.01730 &  -0.00127 &  (3) \\
         &  5259.69265 &  0.00041  &  -12371.5 & 0.00153  &  -0.00321 &  (3) &          &  4376.80135 &  0.00030  &  -17908   & -0.01734 &  -0.00132 &  (5) \\
         &  5927.59126 &  0.00023  &  -9411.5  & 0.00693  &  0.00269  &  (3) &          &  4376.89786 &  0.00036  &  -17907.5 & -0.01883 &  -0.00282 &  (5) \\
         &  5927.70219 &  0.00073  &  -9411    & 0.00504  &  0.00080  &  (3) &          &  5106.65978 &  0.00067  &  -14184.5 & -0.01704 &  -0.00231 &  (3) \\
         &  6536.81308 &  0.00040  &  -6711.5  & 0.00183  &  -0.00157 &  (3) &          &  5106.76188 &  0.00058  &  -14184   & -0.01294 &  0.00179  &  (3) \\
         &  6536.92603 &  0.00044  &  -6711    & 0.00196  &  -0.00144 &  (3) &          &  5657.66619 &  0.00033  &  -11373.5 & -0.00598 &  0.00708  &  (3) \\
         &  6900.99397 &  0.00056  &  -5097.5  & 0.00040  &  -0.00232 &  (1) &          &  6206.69597 &  0.00041  &  -8572.5  & -0.01142 &  -0.00061 &  (3) \\
         &  6901.11007 &  0.00058  &  -5097    & 0.00368  &  0.00096  &  (1) &          &  6206.79169 &  0.00046  &  -8572    & -0.01370 &  -0.00290 &  (3) \\
         &  8051.19109 &  0.00012  &  -0.5     & -0.00034 &  -0.00010 &  (2) &          &  7887.03740 &  0.00017  &  0        & 0.00000  &  0.00016  &  (2) \\
         &  8051.30425 &  0.00012  &  0        & 0.00000  &  0.00024  &  (2) &          &  7887.13498 &  0.00014  &  0.5      & -0.00043 &  -0.00026 &  (2) \\
         &  8135.80195 &  0.00135  &  374.5    & -0.00433 &  -0.00382 &  (1) & J161857  &  3695.65632 &  0.00126  &  -20079   & 0.03921  &  -0.01438 &  (3) \\
         &  8135.91599 &  0.00067  &  375      & -0.00311 &  -0.00260 &  (1) &          &  4259.61430 &  0.00047  &  -17614   & 0.05202  &  0.00126  &  (3) \\
J074350  &  3854.65199 &  0.00055  &  -20357   & -0.03415 &  -0.00345 &  (3) &          &  4259.72781 &  0.00077  &  -17613.5 & 0.05114  &  0.00038  &  (3) \\
         &  3854.77482 &  0.00088  &  -20356.5 & -0.02498 &  0.00572  &  (3) &          &  4571.78330 &  0.00067  &  -16249.5 & 0.04935  &  0.00059  &  (3) \\
         &  4911.60277 &  0.00090  &  -15707.5 & -0.02631 &  0.00419  &  (3) &          &  4571.89900 &  0.00094  &  -16249   & 0.05066  &  0.00190  &  (3) \\
         &  4911.71513 &  0.00061  &  -15707   & -0.02761 &  0.00288  &  (3) &          &  5098.53930 &  0.00134  &  -13947   & 0.03710  &  -0.00757 &  (3) \\
         &  5834.87690 &  0.00045  &  -11646   & -0.02861 &  -0.00160 &  (3) &          &  5098.64910 &  0.00071  &  -13946.5 & 0.03251  &  -0.01216 &  (3) \\
         &  5834.99162 &  0.00050  &  -11645.5 & -0.02755 &  -0.00055 &  (3) &          &  5726.53864 &  0.00072  &  -11202   & 0.03259  &  -0.00602 &  (3) \\
         &  6398.64324 &  0.00075  &  -9166    & -0.02579 &  -0.00244 &  (3) &          &  6119.70278 &  0.00061  &  -9483.5  & 0.03658  &  0.00241  &  (3) \\
         &  7045.84236 &  0.00129  &  -6319    & -0.01809 &  -0.00036 &  (1) &          &  6470.87939 &  0.00059  &  -7948.5  & 0.03436  &  0.00458  &  (3) \\
         &  7045.96023 &  0.00086  &  -6318.5  & -0.01389 &  0.00385  &  (1) &          &  8276.27220 &  0.00035  &  -57      & 0.00191  &  0.00100  &  (2) \\
         &  8031.98624 &  0.00192  &  -1981    & -0.00573 &  0.00053  &  (1) &          &  8289.31081 &  0.00025  &  0        & 0.00000  &  -0.00066 &  (2) \\
         &  8032.10273 &  0.00125  &  -1980.5  & -0.00290 &  0.00336  &  (1) & J224015  &  3831.59328 &  0.00018  &  -20706   & 0.01034  &  0.00068  &  (3) \\
         &  8482.20738 &  0.00020  &  -0.5     & 0.00023  &  0.00008  &  (2) &          &  3831.69837 &  0.00031  &  -20705.5 & 0.00603  &  -0.00363 &  (3) \\
         &  8482.32081 &  0.00020  &  0        & 0.00000  &  -0.00015 &  (2) &          &  4437.67841 &  0.00052  &  -17936   & 0.01393  &  0.00784  &  (3) \\
J104942  &  1491.87662 &  0.00192  &  -29026.5 & 0.02181  &  0.00004  &  (4) &          &  4938.61627 &  0.00032  &  -15646.5 & 0.00461  &  0.00096  &  (3) \\
         &  1491.99182 &  0.00176  &  -29026   & 0.02211  &  0.00034  &  (4) &          &  4938.72031 &  0.00058  &  -15646   & -0.00075 &  -0.00440 &  (3) \\
         &  3674.01032 &  0.00054  &  -19530.5 & -0.01579 &  -0.00705 &  (3) &          &  5624.55348 &  0.00077  &  -12511.5 & -0.00245 &  -0.00351 &  (3) \\
         &  4239.65895 &  0.00076  &  -17069   & -0.01740 &  -0.00430 &  (3) &          &  6322.75066 &  0.00050  &  -9320.5  & -0.00245 &  -0.00178 &  (3) \\
         &  4239.77023 &  0.00106  &  -17068.5 & -0.02102 &  -0.00792 &  (3) &          &  6735.19910 &  0.00102  &  -7435.5  & 0.00422  &  0.00549  &  (1) \\
         &  4939.73994 &  0.00060  &  -14022.5 & -0.01906 &  -0.00258 &  (3) &          &  7422.11993 &  0.00101  &  -4296    & -0.00383 &  -0.00226 &  (1) \\
         &  5718.75189 &  0.00079  &  -10632.5 & -0.02572 &  -0.00812 &  (3) &          &  7422.23278 &  0.00284  &  -4295.5  & -0.00038 &  0.00119  &  (1) \\
         &  6370.70139 &  0.00090  &  -7795.5  & -0.01599 &  0.00043  &  (3) &          &  8158.16863 &  0.00069  &  -932     & -0.00506 &  -0.00414 &  (1) \\
         &  6802.71814 &  0.00042  &  -5915.5  & -0.02136 &  -0.00679 &  (1) &          &  8362.09715 &  0.00021  &  0        & 0.00000  &  0.00056  &  (2) \\
         &  7424.79082 &  0.00123  &  -3208.5  & -0.01457 &  -0.00418 &  (1) &          &  8362.20573 &  0.00030  &  0.5      & -0.00082 &  -0.00026 &  (2) \\
         &  7424.90854 &  0.00077  &  -3208    & -0.01175 &  -0.00136 &  (1) & &&&&&&\\
         &  7788.79668 &  0.00031  &  -1624.5  & -0.01032 &  -0.00319 &  (1) & &&&&&&\\
         &  8162.11548 &  0.00043  &  0        & 0.00000  &  0.00316  &  (2) & &&&&&&\\
         &  8162.23004 &  0.00033  &  0.5      & -0.00034 &  0.00282  &  (2) & &&&&&&\\
         &  8209.91285 &  0.00100  &  208      & -0.00082 &  0.00178  &  (1) & &&&&&&\\
\hline
\end{tabular}
(1) ASAS-SN; (2) This paper; (3) CSS; (4) NSVS; (5) LINEAR.
\end{table*}

\begin{table*}
\footnotesize
\begin{center}
\caption{The derived parameters of the second-order polynomial for the nine targets}
\begin{tabular}{lcccccc}
\hline
Parameters&  $\Delta T_0$ ($\times10^{-4}$ d)& Error &  $\Delta P_0$ ($\times10^{-7}$ d) & Error & $\beta$ ($\times10^{-7}$ d yr$^{-1}$) & Error  \\\hline
J003033   &  -1.05                              & 3.65  &  5.82                                & 3.82  & -1.65                                    & 0.85   \\
J014418   &  -1.32                              & 4.5   &  -3.17                               & 2.09  & 0.74                                     & 0.37   \\
J031700   &  -2.38                              & 2.42  &  -7.06                               & 1.08  & -0.80                                    & 0.21   \\
J074350   &  1.52                               & 5.93  &  3.42                                & 2.75  & 3.01                                     & 0.51   \\
J104942   &  -31.6                              & 6.02  &  26.40                               & 1.53  & 3.83                                     & 0.25   \\
J130945   &  12.0                               & 6.18  & -22.39                               & 2.56  &-1.38                                     & 0.36   \\
J145224   &  -1.64                              & 7.96  &  15.67                               & 2.34  & 1.42                                     & 0.39   \\
J161857   &  6.63                               & 5.13  &  -4.34                               & 1.44  & -2.70                                    & 0.21   \\
J224015   &  -5.61                              & 12.5  &  4.25                                & 4.35  & 1.48                                     & 0.70   \\\hline
\end{tabular}
\end{center}
\end{table*}

\section{Discussions and conclusions}
Based on our new complete light curves of nine USPCBs, we first determined their photometric elements by using the W-D code. We found that all the targets are W-subtype contact binaries. Eight of the targets are shallow contact systems ($f\leq25\%$) while one (J130945) is a medium contact system. Four stars show clearly O'Connell effect, which can be modeled by dark spot or hot spot on one of the two components, and three systems have been detected with third light.
By using all available eclipse timings, we analyzed the period variation behavior of all targets and found that four of them exhibit long-term orbital period decrease, while five of them display long-term orbital period increase.

\subsection{The reliability of the mass ratio}
Mass ratio is a very important parameter for contact binaries. Due to the statistical study of Pribulla et al. (2003) and the numerical simulations of Terrell \& Wilson (2005), the photometric mass ratio should be reliable for totally eclipsing
contact binaries. According to our study, the orbital inclinations of J003033, J014418, and J074350 are greater than 80$^\circ$, meaning they are total eclipse binaries, hence their photometric mass ratios are reliable. Recently, Zhang et al. (2017b) proposed two criteria for the reliability of the light-curve synthetics. One is the degree of symmetry of the light curve, the more symmetric the light curve, the more reliable the solution result. The other is the degree of sharpness of the q-search curve bottom, the sharper the bottom, the more reliable the photometric mass ratio. To discuss the reliability of the photometric mass ratios of our nine targets, we can discuss the q-search curve presented in Figure 2 for each star. The q-search curves of J003033, J014418, J074350, J104942, J145224, J161857, and J224015 show very sharp bottom, the position of the minimum can be easily discovered, indicating the uncertainties of their photometric mass ratios are very small. Quantitatively, the uncertainties should be less than 0.2. However, the q-search curves of J031700 and J130945 show very shallow minima (a very wide bottom, visually, the width is about 0.5 (from 3.9 to 4.4 for J031700 and from 3.5 to 4.0 for J130945), indicating the uncertainties of their photometric mass ratios are very large. Quantitatively, the uncertainties may be no less than 0.5.
According to our discussions, the mass ratios of J003033, J014418, and J074350 are the most reliable ones because the three binaries are not only totally eclipsing contact binaries but also manifest very sharp q-search curve bottom, the mass ratios of J104942, J145224, J161857, and J224015 are the second reliable ones because they only meet the criterion suggested by Zhang et al. (2017b), and the mass ratios of J031700 and J130945 are likely to be unreliable and should be treated as preliminary results.

\subsection{The secular orbital period changes}
All the nine systems manifest secular orbital period variations, continuous decreasing or increasing. To further discuss the reason of the long-term orbital period changes, we need to know in advance the absolute parameters of these systems. Unfortunately, the absolute parameters of the systems are not possible since their radial velocity curves are not yet available. Nonetheless, we made a preliminary estimation of the parameters for the systems, by using the adopted light curve parameters and some well-established relations. According to Hilditch et al. (1988), the more massive component of a contact binary is normally a main-sequence star. So, assuming that the more massive components of our nine targets are main-sequence stars, we estimated the masses by interpolating from the online Table\nolinebreak\footnotemark[1]\footnotetext[4]{http://www.pas.rochester.edu/\textasciitilde emamajek/EEM\_dwarf\_UBVIJHK\_colors\_Teff.txt} provided by Pecaut \& Mamajek (2013) due to their temperatures. Based on our light curve solutions, their absolute parameters were derived as shown in Table 8. Secular orbital period increase is usually caused by the mass transfer from the less massive component to the more massive one, while continuous orbital period decrease generally results from the mass transfer from the more massive component to the less massive one, from the AML or their combination. Using the following equation (Kwee 1958),
\begin{eqnarray}
{\dot{P}\over P}=-3\dot{M_1}({1\over M_1}-{1\over M_2}) .
\end{eqnarray}
we calculated the mass transfer rate for all systems as displayed in Table 8. For J014418, J074350, J104942, J145224, and J224015, the less massive component is losing mass. For J003033, J031700, J130945, and J161857, the less massive component is receiving mass, we computed the thermal timescale of the more massive components of the four stars using $\tau_{th}={GM^2\over RL}$ and determined $\tau_{th}=1.53\times10^8$ years for J003033, $\tau_{th}=1.08\times10^8$ years for J031700, $\tau_{th}=1.24\times10^8$ years for J130945, and $\tau_{th}=1.87\times10^8$ years for J161857. The thermal mass transfer rate was calculated to be $M_2/\tau_{th}=0.51\times10^{-8}\,M_\odot$ yr$^{-1}$ for J003033, $M_2/\tau_{th}=0.69\times10^{-8}\,M_\odot$ yr$^{-1}$ for J031700, $M_2/\tau_{th}=0.38\times10^{-8}\,M_\odot$ yr$^{-1}$ for J130945, and $M_1/\tau_{th}=0.36\times10^{-8}\,M_\odot$ yr$^{-1}$ for J161857. These four values are quite different from those determined by Equation (4), revealing that the long-term period decrease of the four stars cannot be caused by mass transfer but results from AML. For the long-term orbital period variations of the nine stars, we cannot exclude the possibility of very long period cyclic variations because the time span of the eclipsing times of every system is less than twenty years. Therefore, the reason for the observed period variations is still debated, and continuous observations of these targets are needed in the future.

\begin{table*}
\footnotesize
\begin{center}
\caption{Absolute parameters of the stars}
\begin{tabular}{lcccccccc}
\hline
Paramers& M$_1$($M_\odot$)&M$_2$($M_\odot$) &a($R_\odot$)&R$_1$($R_\odot$)& R$_2$($R_\odot$)&L$_1$($L_\odot$) & L$_2$($L_\odot$)& dM$_1$/dt ($10^{-1}$ $M_\odot$ yr$^{-1}$)\\\hline
J003033 &  0.38  &0.79   &1.65  & 0.55 	& 0.75 	& 0.10 	&0.17 & 1.80($\pm$0.92)    \\
J014418 &  0.34  &0.72   &1.55  & 0.51 	& 0.70 	& 0.13 	&0.21 &-0.74($\pm$0.37)   \\
J031700 &  0.19  &0.75   &1.53  & 0.41 	& 0.77 	& 0.07 	&0.21 & 0.31($\pm$0.08)   \\
J074350 &  0.28  &0.66   &1.54  & 0.48 	& 0.72 	& 0.10 	&0.20 &-2.19($\pm$0.37)   \\
J104942 &  0.48  &0.69   &1.66  & 0.60 	& 0.70 	& 0.19 	&0.23 &-8.56($\pm$0.56)   \\
J133417 &  0.13  &0.47   &1.26  & 0.37 	& 0.63 	& 0.04 	&0.09 & 0.39($\pm$0.10)    \\
J145224 &  0.28  &0.54   &1.33  & 0.44 	& 0.61 	& 0.04 	&0.08 &-1.40($\pm$0.38)   \\
J161857 &  0.68  &0.40   &1.62  & 0.69 	& 0.54 	& 0.23 	&0.14 &-3.95($\pm$0.31)   \\
J224015 &  0.28  &0.69   &1.51  & 0.47 	& 0.70 	& 0.09 	&0.17 &-1.03($\pm$0.49)   \\\hline
\end{tabular}
\end{center}
\end{table*}

\subsection{Statistics on well-studied USPCBs with orbital period variations }
The statistic of well-studied USPCBs with orbital period variations was carried out. With literature search in combination with our results, we collected data of the 19 systems shown in Table 9. Our present research has thus doubled the number of this kind of systems. Among the 19 systems, 7 systems display secular orbital period decrease (3 of them also show cyclic variation), 10 systems show continuous orbital period increase, and 2 systems do not exhibit long-term period changes but manifest cyclic variations. For the systems with increasing orbital period, it is expected that the separation between their two components will increase, and this will lead to the decrease of the degree of contact. The shallow contact systems will evolve from the present weak-contact configuration to a broken-contact configuration due to the thermal relaxation
oscillation theory (e.g., Lucy 1976; Flannery 1976; Robertson \& Eggleton 1977; Lucy \& Wilson 1979; Qian 2001). On the other hand, the seven systems with decreasing period will evolve from the present shallow or medium contact configuration to a deep contact configuration.
Due to our statistic, most of the systems are shallow contact binaries, meaning that they are newly formed contact systems. This can support the long timescale AML theory suggested by Stepien (2006, 2011). However, there are three deep contact systems (1SWASP J075102.16+342405.3, 1SWASP J234401.81$-$212229.1, and 1SWASP J074658.62+224448.5). As suggested by Li et al. (2019), the long timescale AML theory can not explain their existence and an additional companion may play a very import role during their formation. The orbital period investigation (Lohr et al. 2013a; Koen 2014) of 1SWASP J234401.81$-$212229.1 has proved that it is a hierarchical multiple-star system. In order to check whether 1SWASP J074658.62+224448.5 and 1SWASP J075102.16+342405.3 exhibit cyclic variation in the orbital period, we reanalyzed their $O-C$ diagrams by using all the available eclipsing times including the data of NSVS, SuperWASP, CSS, ASAS-SN, Jiang et al. (2015a, 2015b), and Kjurkchieva et al. (2018). Using the same linear ephemerides those given by Jiang et al. (2015a, 2015b), we calculated the $O-C$ values of the two systems and constructed the $O-C$ diagrams displayed in Figure 6. As seen in this figure, the $O-C$ curves of the two stars exhibit a combination of an upward parabola and a periodic variation. Therefore, a quadratic term plus the light travel time effect term taken from Irwin(1952) were applied to fit their curves,
\begin{eqnarray}
O-C&=& \Delta T_{0} + \Delta P_{0}\times E+{\beta \over 2}\times E^2 + K{1\over\sqrt{1-e^2\cos^2\omega} }[(1-e^{2})\frac{\sin(\nu+\omega)}{1+e\cos\nu}+e\sin\omega].
\end{eqnarray}
The determined parameters are shown in Table 10. According to the light curve study of Jiang et al. (2015a, 2015b), non-ignorable third light was detected for the two stars. In addition, due to the statistic study of Liao \& Qian (2010), the cyclic oscillation detected in the $O-C$ diagram was most likely caused by a third companion. Therefore, for these three deep contact USPCBs, a third body plays a crucial role during their formation and evolution by removing the angular momentum in order to decrease the AML timescale.

\renewcommand\arraystretch{1.3}
\begin{table*}
\tiny
\begin{center}
\caption{Well-studied USPCBs with orbital period changes}
\begin{tabular}{p{3.5cm}p{0.8cm}p{0.7cm}p{0.7cm}p{0.3cm}p{0.5cm}p{0.5cm}p{0.5cm}p{1.7cm}p{0.5cm}p{0.9cm}p{1cm}}
\hline
Name                        &  Period   &Subtype & $q$      & $i$   &  $f$  & $T_p$& $T_s$ & dP/dt             & T$_3$      & Amplitude       & Ref.     \\\hline
                            &   d       &        & $M_s/M_p$&       &  \%   &  K   &  K    & 10$^{-7}$ d yr$^{-1}$ &  yr    &  d              &          \\\hline
CRTS J145224.5+011522       &  0.196014 &  W     &  0.517   & 77.0  &  14   & 3807 & 3898  & 1.42              & $-$        & $-$             & (1)      \\
SDSS J001641$-$000925       &  0.198563 &  A     &  0.620   & 53.3  &  22   & 4342 & 3889  & $-$               & 5.69       & 0.00255         & (2); (3) \\
1SWASP J075102.16+342405.3  &  0.209172 &  A     &  0.780   & 77.0  &  96   & 3950 & 3876  & 1.35              & $-$        & $-$             & (4)      \\
CRTS J130945.0+371627       &  0.211132 &  W     &  0.273   & 73.1  &  29   & 3654 & 3762  & $-$1.38           & $-$        & $-$             & (1)      \\
1SWASP J234401.81$-$212229.1&  0.213676 &  A     &  0.422   & 79.4  &  deep?& 4400 & 4400  & $-$               & 4.19       & 0.0073          & (5); (6) \\
CRTS J014418.3+190625       &  0.217372 &  W     &  0.475   & 82.2  &  15   & 4607 & 4819  & 0.74              & $-$        & $-$             & (1)      \\
CRTS J224015.4+184738       &  0.218802 &  W     &  0.400   & 77.3  &  14   & 4441 & 4643  & 1.48              & $-$        & $-$             & (1)      \\
CC Com                      &  0.220686 &  W     &  0.526   & 89.8  &  17   & 4200 & 4300  & $-$0.20           & 23.6       & 0.0028          & (7); (8) \\
1SWASP J074658.62+224448.5  &  0.220850 &  W     &  0.365   & 81.7  &  51   & 4543 & 4717  & 5.39              & $-$        & $-$             & (9); (10)\\
1SWASP J140533.33+114639.1  &  0.225126 &  W     &  0.646   & 68.6  &  8    & 4523 & 4680  & 2.09              & $-$        & $-$             & (11)     \\
1SWASP J031700.67+190839.6  &  0.225652 &  W     &  0.258   & 73.7  &  4    & 4791 & 4968  &$-$0.80            & $-$        & $-$             & (1)      \\
1SWASP J160156.04+202821.6  &  0.226529 &  A     &  0.670   & 79.5  &  10   & 4500 & 4500  & 10.88             & $-$        & $-$             & (12)     \\
1SWASP J003033.05+574347.6  &  0.226618 &  W     &  0.484   & 82.8  &  24   & 5067 & 5246  & $-$1.65           & $-$        & $-$             & (1)      \\
CRTS J074350.9+451620       &  0.227324 &  W     &  0.430   & 81.1  &  18   & 4312 & 4471  & 3.01              & $-$        & $-$             & (1)      \\
V1104 Her$^a$               &  0.227876 &  W     &  0.623   & 83.4  &  15   & 3902 & 4050  & $-$0.18           & 8.28       & 0.00125         & (13)     \\
1SWASP J161857.80+261338.9  &  0.228781 &  W     &  0.597   & 50.8  &  4    & 4400 & 4578  & $-$2.70           & $-$        & $-$             & (1)      \\
1SWASP J200503.05$-$343726.5&  0.228884 &  W     &  0.934   & 73.8  &  9    & 4270 & 4500  & 0.54              & $-$        & $-$             & (14)     \\
1SWASP J161335.80$-$284722.2&  0.229778 &  W     &  0.909   & 78.0  &  19   & 4008 & 4209  & $-$4.26           & 4.79       & 0.00196         & (15)     \\
1SWASP J104942.44+141021.5  &  0.229799 &  W     &  0.690   & 59.6  &  15   & 4457 & 4560  & 3.83              & $-$        & $-$             & (1)      \\\hline
\end{tabular}
\end{center}
(1) This paper; (2) Davenport  et al. 2013; (3) Qian et al. 2015; (4) Jiang et al. 2015a; (5) Lohr et al. 2013a; (6) Koen 2014; (7) Yang et al. 2009;
(8) K\"{o}se et al. 2011; (9) Jiang et al. 2015b; (10) Kjurkchieva et al. 2018; (11) Zhang et al. 2018; (12) Lohr et al. 2014; (13) Liu et al. 2015;
(14) Zhang et al. 2017a; (15) Fang et al. 2019.\\
$^a$ We only show the longer period cyclic variation.\\
Note$-$ "p" means the more mass component, while "s" represents the less massive one.
\end{table*}

\begin{figure}\centering
\includegraphics[width=0.45\textwidth]{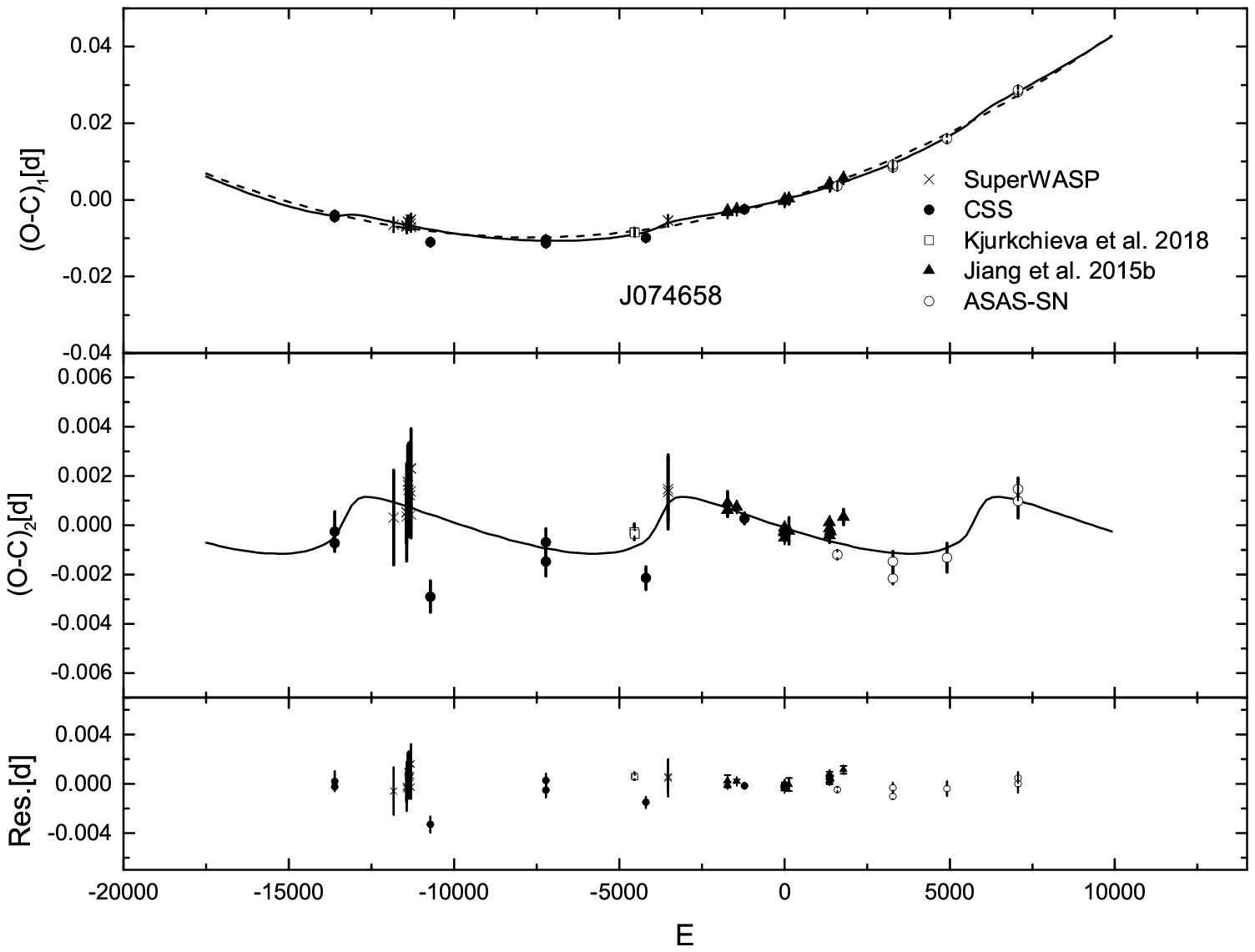}
\includegraphics[width=0.45\textwidth]{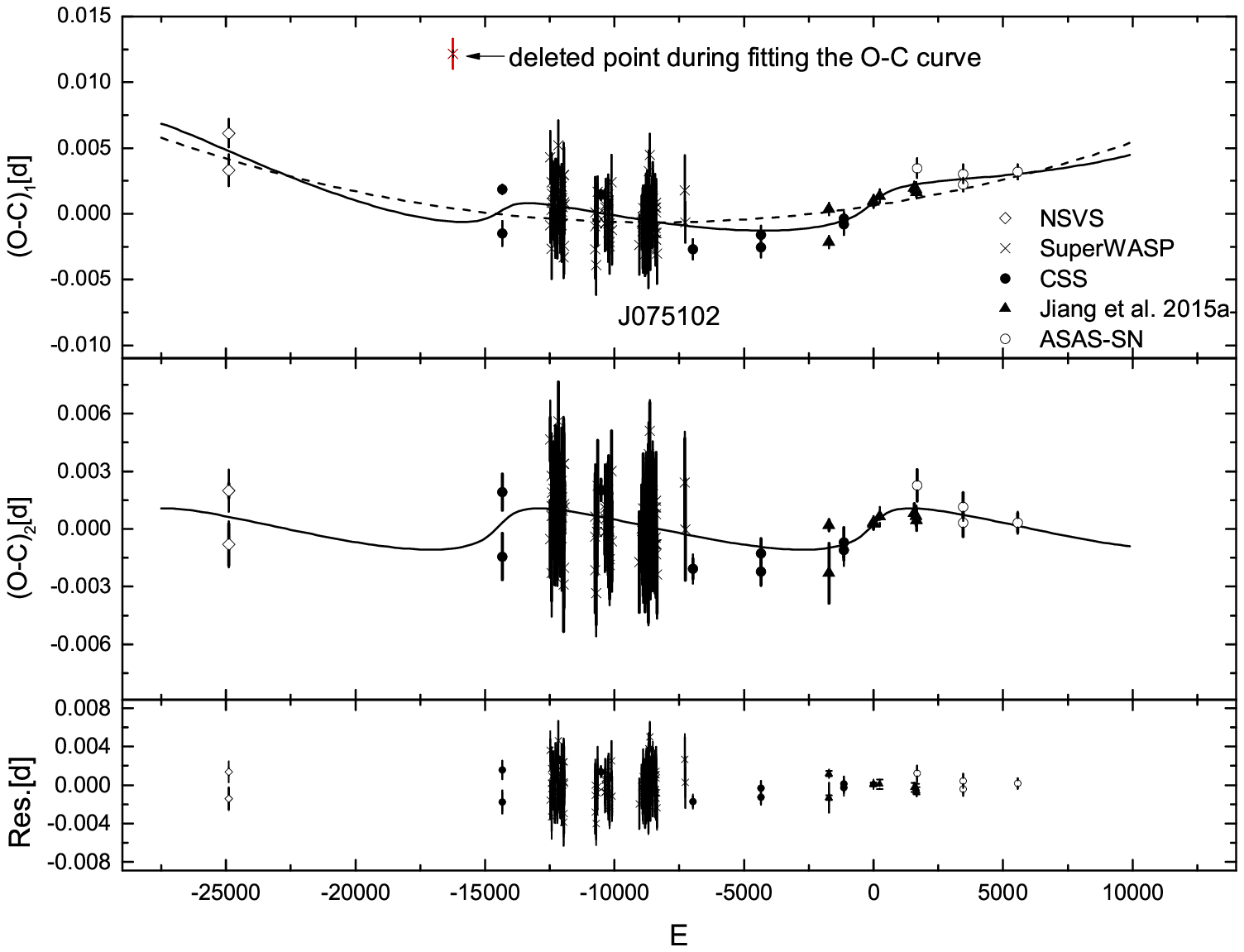}

\caption{The $O-C$ diagrams for 1SWASP J074658.62+224448.5 and 1SWASP J075102.16+342405.3. }
\end{figure}

\begin{table}
\begin{center}
\caption{Parameters determined by Equations (5) for 1SWASP J074658.62+224448.5 and 1SWASP J075102.16+342405.3}
\begin{tabular}{lcc}
\hline\hline
Parameters &     1SWASP J074658.62+224448.5 &     1SWASP J075102.16+342405.3 \\
\hline
$\Delta T_0$ (days) &    0.000228$(\pm0.000282)$ &  $0.000632(\pm0.000550)$  \\

$\Delta P_0$ (days) & $2.62(\pm0.06)\times10^{-6}$ & $3.03(\pm0.63)\times10^{-7}$  \\

$\beta$ (days $yr^{-1}$) & $11.33(\pm0.39)\times10^{-7}$ & $1.24(\pm0.24)\times10^{-7}$   \\

$K$ (d) &     0.00178($\pm0.00099$) & 0.00145($\pm0.00086$)    \\

$e$ &       0.80($\pm0.43$) &   0.69($\pm0.36$)   \\

$P_{3}$ (yr) &     5.77($\pm0.39$) &   8.24($\pm0.58$)   \\

$\omega (^\circ)$ &       17.8($ \pm13.1$) &  12.1($ \pm11.6$)   \\

$ T_P$ (HJD) &  2453570($\pm836$) &  2456668($\pm997$)    \\
\hline
\end{tabular}
\end{center}
\end{table}

In conclusion, we have analyzed the light curves and orbital period variations of nine USPCBs and have done statistical work on well-studied USPCBs with orbital period variations. We found that all our targets are W-subtype contact systems. One of them is a medium contact binary while the others are shallow contact ones. 19 systems with orbital period variations were identified. Seven of them display long-term decrease (three of them also exhibit cyclic variations), ten of them manifest long-term increase while two of them only show cyclic variation. Since the short period end of contact binaries is still a mystery, more USPCBs should be observed and studied in the future.

\acknowledgments{This work is supported by National Natural Science Foundation of China (NSFC) (No. 11703016), and by the Joint Research Fund in Astronomy (No. U1931103) under cooperative agreement between NSFC and Chinese Academy of Sciences (CAS), and by the Natural Science Foundation of Shandong Province (Nos. ZR2014AQ019, ZR2017PA009, ZR2017PA010, JQ201702), and by Young Scholars Program of Shandong University, Weihai (Nos. 20820162003, 20820171006), and by the Open Research Program of Key Laboratory for the Structure and Evolution of Celestial Objects (No. OP201704). Work by CHK was supported by the grant of National Research Foundation of Korea (2020R1A4A2002885). RM acknowledge the financial support from the UNAM under DGAPA grant PAPIIT IN 100918. Thanks Dr. Lohr very much for sending us eclipsing times and uncertainties of the targets observed by SuperWASP. Many thanks to the anonymous referee for his/her very valuable and helpful recommendations.

This work is partly supported by the Supercomputing Center of Shandong University, Weihai.

We acknowledge the support of the staff of the Xinglong 85cm
telescope, NEXT, SPM and WHOT. This work was partially supported by the Open Project Program of the Key
Laboratory of Optical Astronomy, National Astronomical Observatories, Chinese
Academy of Sciences.

This publication makes use of data products from the AAVSO
Photometric All Sky Survey (APASS). Funded by the Robert Martin Ayers
Sciences Fund and the National Science Foundation.

This publication makes use of data products from the Two Micron All Sky Survey, which is a joint project of the University of Massachusetts and the Infrared Processing and Analysis Center/California Institute of Technology, funded by the National Aeronautics and Space Administration and the National Science Foundation.

This paper makes use of data from the DR1 of the WASP data (\citealt{Butters2010}) as provided by the WASP consortium,
and the computing and storage facilities at the CERIT Scientific Cloud, reg. no. CZ.1.05/3.2.00/08.0144
which is operated by Masaryk University, Czech Republic.}

Funding for SDSS-III has been provided by the Alfred P. Sloan Foundation, the Participating Institutions, the National Science
Foundation, and the U.S. Department of Energy Office of Science. The
SDSS-III web site is http://www.sdss3.org/.

SDSS-III is managed by the Astrophysical Research Consortium for the
Participating Institutions of the SDSS-III Collaboration including the
University of Arizona, the Brazilian Participation Group, Brookhaven
National Laboratory, University of Cambridge, University of Florida,
the French Participation Group, the German Participation Group, the
Instituto de Astrofisica de Canarias, the Michigan State/Notre
Dame/JINA Participation Group, Johns Hopkins University, Lawrence
Berkeley National Laboratory, Max Planck Institute for Astrophysics,
New Mexico State University, New York University, Ohio State
University, Pennsylvania State University, University of Portsmouth,
Princeton University, the Spanish Participation Group, University of
Tokyo, University of Utah, Vanderbilt University, University of
Virginia, University of Washington, and Yale University.

\end{document}